\documentclass[notitlepage,a4,10.5pt]{article}

\usepackage[colorlinks=true,citecolor=blue,linkcolor=blue]{hyperref}

\usepackage[left=2.5cm,right=2.5cm,top=3cm,bottom=3cm]{geometry} 

\usepackage{tikz}
\usetikzlibrary{arrows.meta,positioning,calc,shapes,fit}
\usepackage{subcaption} 

\usepackage{booktabs}

\usepackage{tabularx}
\usepackage{array}
\newcolumntype{Y}{>{\raggedright\arraybackslash}X}
\usepackage{booktabs}  
\usepackage{multirow}  

\usepackage{graphicx}

\usepackage{amsmath}%
\numberwithin{equation}{section}

\usepackage{amsfonts}%
\usepackage{amssymb}%
\usepackage{mathrsfs}
\usepackage{amsthm}
\usepackage{float}

\usepackage{authblk}
\usepackage{enumitem}

\usepackage{xcolor}
\usepackage{appendix}

\newcommand{\ml}{\mathcal}

\newcommand{\Norm}[1]{\left\lvert\left\lvert #1 \right\rvert\right\rvert}

\newcommand{\bol}{\boldsymbol}

\newcommand{\abs}[1]{\left\lvert{#1}\right\rvert}

\newcommand{\lr}[1]{\left({#1}\right)}
\newcommand{\lrs}[1]{\left[{#1}\right]}
\newcommand{\lrc}[1]{\left\{{#1}\right\}}

\newcommand{\p}{\partial}

\newtheorem{thm}{{Theorem}}[section]

\theoremstyle{definition}
\newtheorem{mydef}[thm]{Definition}

\theoremstyle{remark}
\newtheorem{remark}[thm]{\textbf{Remark}}

\newtheorem{q}[thm]{\textbf{Question}}

\newcommand{\eq}[1]{\begin{equation}\begin{split}{#1}\end{split}\end{equation}}
\newcommand{\sys}[2]{\begin{subequations}\begin{align}{#1}\end{align}\label{#2}\end{subequations}}

\definecolor{myyellow}{RGB}{255,255,180}

\begin{document}

\title{
{
Compact Core--Shell Equilibria in Gravitational Vlasov--Poisson Systems with Positive and Negative Mass 
}
}
\author{Naoki Sato\thanks{Corresponding author} {}\thanks{National Institute for Fusion Science,  322-6 Oroshi-cho Toki-city, Gifu 509-5292, Japan,\ Email: \href{sato.naoki@nifs.ac.jp}{sato.naoki@nifs.ac.jp}
}  
{}\thanks{Graduate School of Frontier Sciences, The University of Tokyo,  
Kashiwa, Chiba 277-8561, Japan}
}
 
\date{\today}
\setcounter{Maxaffil}{0}
\renewcommand\Affilfont{\itshape\small}

    \maketitle

    \begin{abstract}
We construct regular, spherically symmetric, compact energy-cutoff equilibria for gravitational Vlasov--Poisson systems sourced by positive  and negative mass distributions in the context of stellar dynamics. Motivated by Bondi's notion of negative mass and its relation to the weak field limit of Einstein gravity with a signed mass--energy source, we show that the internal structure of the steady states depends qualitatively on the type of negative mass present. In the Bondi convention, the signs of inertial, passive gravitational, and active gravitational mass are reversed together, preserving the equivalence principle. The resulting equilibria consist of a central overlap core containing both mass species, surrounded by a finite positive mass shell and an exterior vacuum. Thus, although a pure Bondi negative mass gas cannot form a compact steady state, Bondi negative mass can be spatially confined by a suitable positive mass distribution. In the gravitational-charge convention, only the passive and active gravitational masses change sign, so that the two species obey opposite free-fall laws and unlike masses repel. The resulting equilibria are spatially segregated into a negative mass core, a vacuum gap, a positive mass shell, and an exterior vacuum. For cutoff exponent \(n=-1/2\), the radial matching is explicit, while for the regular cutoff \(n=1/2\) the matter regions satisfy Lane--Emden-type equations. These results provide a kinetic framework for investigating the distribution of negative mass in astrophysical and cosmological settings. 
\end{abstract}

\tableofcontents

\section{Introduction}

\subsection{Motivation}
Negative mass distributions arise as solutions of the Einstein field equations
\cite{Bondi1957,Bonnor1989}, and 
their potential cosmological relevance has been explored in several theoretical studies
\cite{Benoit2012,Farnes2018,Manfredi2018,Chardin2018,Manfredi2026}.
Although at present there is no experimental evidence for negative mass, we may consider the following question, independently of its formation history or observability: 
\begin{q}\label{q1}
If negative mass were present, could it form a compact self-consistent
distribution, and where would it be relative to positive mass?
\end{q}
\noindent
In this work, we address Question~\ref{q1} within classical 
Vlasov--Poisson kinetic theory and show that the answer depends crucially on the type of negative mass considered.

The gravitational Vlasov--Poisson system is a standard mean-field model
for collisionless matter in stellar dynamics and plasma physics
\cite{Glassey1996,Rein2007,BinneyTremaine2008}. In this description,
particles interact through a self-consistent gravitational or electrostatic
field \cite{Marsden1984}. Distribution functions depending on the
single particle energy provide an important class of spherically symmetric
equilibria, including compactly supported models of finite self-gravitating
systems
\cite{AnEvansSanders2017,PostiEtAl2015,TaruyaSakagami2005}. The Hamiltonian and noncanonical Poisson
structure of the Vlasov equations was established in
\cite{Morrison1980MaxwellVlasov,MarsdenWeinstein1982}. The associated
energy--Casimir method and its use in stability theory are classical in
fluid and plasma mechanics \cite{HolmMarsdenRatiuWeinstein1985}, and
have been developed rigorously for the gravitational Vlasov--Poisson
system \cite{Rein1994EnergyCasimir,GuoRein1999,BattMorrisonRein1995}.

The occurrence of negative mass and negative energy is not limited to cosmology. Negative matter has motivated proposals for gravity-based propulsion \cite{Forward1990,Landis1991}. Negative energy densities are characteristic of warp-drive geometries \cite{Alcubierre1994,Natario2002,Everett1996}. More recent work has explored warp-drive configurations with positive energy densities or reduced exotic-matter requirements \cite{Lentz2021,Bobrick2021}, while general analyses indicate that physically reasonable warp drives nevertheless violate the null energy condition \cite{Santiago2022}. Negative mass configurations have also been constructed as bubbles in de Sitter spacetime \cite{Belletete2013,Mbarek2014,Johnson2020}, while traversable wormhole geometries involve exotic stress--energy whose negative energy content is constrained by quantum field theory \cite{MorrisThorne1988,MorrisThorneYurtsever1988,FordRoman1996}. Antimatter is distinct from negative mass, but its potentially anomalous gravitational behavior has motivated related investigations \cite{Schiff1959,Good1961,Nieto1991}.

These systems are physically distinct from the gravitational Vlasov--Poisson equation  studied here and do not constitute evidence for negative mass. Rather, they illustrate the broader theoretical occurrence of negative mass and negative energy in gravitational physics, motivating a systematic investigation of negative mass distributions and their equilibria within classical collisionless kinetic theory.

\subsection{Negative mass in classical mechanics}\label{subsec:negative mass-classical}

In Newtonian gravity, there are three notions of mass: inertial mass $m^I$, which measures resistance to acceleration, active gravitational mass $m^A$, which sources the gravitational field, and passive gravitational mass $m^P$, which determines the response to a gravitational field. For ordinary matter, all three masses are identical and positive. Different choices of their signs, however, lead to different dynamics and,  as we show in this work, to qualitatively different equilibrium structures.

In this work, we examine two types of negative mass (see Table~\ref{tab:mass-signs}). 
\begin{table}[t]
\centering
\renewcommand{\arraystretch}{1.25}
\begin{tabular}{c|ccc|ccc}
\toprule
& \multicolumn{3}{c|}{Positive species}
& \multicolumn{3}{c}{Negative species}
\\
Mass type
& \(m_+^I\) & \(m_+^P\) & \(m_+^A\)
& \(m_-^I\) & \(m_-^P\) & \(m_-^A\)
\\
\midrule
Bondi
& \(m\) & \(m\) & \(m\)
& \(-m\) & \(-m\) & \(-m\)
\\
Gravitational charge
& \(m\) & \(m\) & \(m\)
& \(m\) & \(-m\) & \(-m\)
\\
\bottomrule
\end{tabular}
\caption{
Inertial, passive gravitational, and active gravitational masses in the two
negative mass conventions considered in this paper. Here \(m>0\). The Bondi
choice preserves \(m^I=m^P\) for both species, whereas the
gravitational-charge choice reverses the sign of \(m^P/m^I\) for the negative
species.
}
\label{tab:mass-signs}
\end{table} 
The first corresponds to Bondi's notion of negative mass
\cite{Bondi1957}, in which the signs of inertial, active gravitational, and passive gravitational mass are reversed simultaneously. Since the identity $m^I=m^P$ is preserved, Bondi negative mass remains consistent with the equivalence principle. In particular, positive and negative masses obey the same free-fall law. At the kinetic level, positive and negative mass distributions are therefore transported by the same gravitational potential field, while sourcing Poisson's equation with opposite signs.

The second type considered here is the gravitational-charge convention, in which the negative species has positive inertial mass but negative passive and active gravitational masses. This choice violates the usual equivalence principle, since $m^P\neq m^I$ for the negative species, but provides a useful comparison model because positive and negative masses respond oppositely to the same gravitational potential.
Specefically, we will see that the Bondi and gravitational-charge models have the same source term in
the Poisson equation but different Vlasov transport equations. Their
comparison therefore separates the effect of the source sign from that
of the dynamical response to the gravitational field.

At the particle level, the behavior of Bondi negative mass and gravitational-charge negative mass can be characterized by examining the two-body acceleration vectors in one spatial dimension.
Let \(\bol{x}_1\) and \(\bol{x}_2\) denote the spatial positions of two point particles. The
gravitational potential produced by particle \(j\) is
\(\Phi_j(\bol{x})=-Gm_j^A/\abs{\bol{x}-\bol{x}_j}\), where $G$ is the gravitational constant, 
and particle \(i\)
satisfies
\eq{
m_i^I\ddot{\bol{x}}_i
=
-m_i^P\nabla\Phi_j(\bol{x}_i).
}
Therefore
\eq{
\ddot{\bol{x}}_1
=
G\frac{m_1^Pm_2^A}{m_1^I}
\frac{\bol{x}_2-\bol{x}_1}{\abs{\bol{x}_1-\bol{x}_2}^3},
\qquad
\ddot{\bol{x}}_2
=
G\frac{m_2^Pm_1^A}{m_2^I}
\frac{\bol{x}_1-\bol{x}_2}{\abs{\bol{x}_1-\bol{x}_2}^3}.\label{eq:point-mass-acceleration}
}
Note that the acceleration vector  depends on the sign of the active
mass sourcing the potential, and on the ratio \(m^P/m^I\), which determines 
the response to that potential. 
The acceleration patterns obtained from
\eqref{eq:point-mass-acceleration} are summarized in
Table~\ref{tab:two-body-accelerations}.

\begin{table}[t]
\centering
\renewcommand{\arraystretch}{1.35}
\begin{tabular}{c|c|c|c}
\toprule
Mass type & Pair & Acceleration direction & Behavior \\
\midrule
& \(++\)
& \(+\!\longrightarrow \qquad \longleftarrow\!+\)
& mutual attraction
\\
Bondi
& \(+-\)
& \(\longleftarrow\!+ \qquad \longleftarrow\!-\)
& runaway pair
\\
& \(--\)
& \(\longleftarrow\!- \qquad -\!\longrightarrow\)
& mutual repulsion
\\
\midrule
& \(++\)
& \(+\!\longrightarrow \qquad \longleftarrow\!+\)
& mutual attraction
\\
Gravitational charge
& \(+-\)
& \(\longleftarrow\!+ \qquad -\!\longrightarrow\)
& mutual repulsion
\\
& \(--\)
& \(-\!\longrightarrow \qquad \longleftarrow\!-\)
& mutual attraction
\\
\bottomrule
\end{tabular}
\caption{
Two-body acceleration directions for the Bondi and gravitational-charge sign
conventions. 
Arrows indicate  acceleration vectors. The \(+-\) Bondi
case gives a runaway pair where both particles accelerate in the
same direction. In the gravitational-charge convention, like species attract
and unlike species repel.
}
\label{tab:two-body-accelerations}
\end{table}

\subsection{Gravitational Vlasov-Poisson system with positive and negative mass}\label{sec:model}
Throughout the paper, the spatial domain is the Euclidean space
\(\mathbb{R}^3\), and the phase space coordinates are  $\lr{\bol{x},\bol{v}}\in\mathbb{R}^6$, with $\bol{x}$ and $\bol{v}$ the particle position and velocity respectively.  
Let $t\in[0,+\infty)$ denote the time variable and 
\(f_{\pm}\lr{\bol{x},\bol{v},t}\) the phase-space distribution
functions of the positive mass and negative mass particle species. In a
collisionless Vlasov--Poisson regime, conservation of particle number in phase
space is expressed by the Liouville equations
\eq{
\frac{\p f_{\pm}}{\p t}
+\bol{v}\cdot\frac{\p f_{\pm}}{\p\bol{x}}
-\frac{m_{\pm}^{P}}{m_{\pm}^{I}}\nabla\Phi\cdot\frac{\p f_{\pm}}{\p\bol{v}}
=0.\label{LVP}
}
Here \(m_\pm^I\), \(m_\pm^P\), and \(m_\pm^A\) denote the inertial, passive
gravitational, and active gravitational masses of each species, respectively. 
The gravitational potential per unit passive mass, \(\Phi\lr{\bol{x},t}\),
is defined so that \(m_\pm^P\Phi\lr{\bol{x},t}\) is the mean-field
gravitational potential energy of a particle of species \(\pm\), namely
\eq{
m_{\pm}^P\Phi\lr{\bol{x},t}
=
-\int_{\mathbb{R}^6}
G\frac{m_{\pm}^P}{\abs{\bol{x}-\bol{x}'}}
\lr{
m_{+}^Af_{+}\lr{\bol{x}',\bol{v}',t}
+
m_{-}^Af_{-}\lr{\bol{x}',\bol{v}',t}
}
\,d\bol{x}'d\bol{v}'.
}
Recalling the distributional identity 
\(\Delta(1/\abs{\bol{x}-\bol{x}'})=-4\pi\delta\lr{\bol{x}-\bol{x}'}\),
where \(\Delta\) denotes the Laplacian operator in the \(\bol{x}\)
variables, the gravitational potential \(\Phi\) is determined by the Poisson
equation
\eq{
\Delta\Phi
=
4\pi G\lr{m^A_{+}\rho_{+}+m^A_{-}\rho_{-}},
}
where \(\rho_{\pm}\) are the number densities,
\eq{
\rho_{\pm}\lr{\bol{x},t}
=
\int_{\mathbb{R}^3}
f_{\pm}\lr{\bol{x},\bol{v},t}\,d\bol{v}.
}
Hence, the complete set of governing equations can be written as follows: 



\begin{mydef}[{Vlasov--Poisson system}]
The gravitational Vlasov--Poisson system with positive and negative mass is 
\sys{
&\frac{\p f_{+}}{\p t}
+\bol{v}\cdot\frac{\p f_{+}}{\p \bol{x}}
-\frac{m_{+}^P}{m^I_{+}}\nabla\Phi\cdot\frac{\p f_{+}}{\p\bol{v}}=0,\\
&\frac{\p f_{-}}{\p t}
+\bol{v}\cdot\frac{\p f_{-}}{\p \bol{x}}
-\frac{m^P_{-}}{m^I_-}\nabla\Phi\cdot\frac{\p f_{-}}{\p\bol{v}}=0,\\
&\Delta\Phi=4\pi G \lr{m_+^A\rho_++m_-^A\rho_-}.\label{VPP}
}{VP}
\end{mydef} 

We note that the mixed-sign gravitational Vlasov--Poisson system \eqref{VP} has been previously formulated for different choices of active, passive, and inertial mass in  \cite{Manfredi2018}. There, the Bondi and ``antiplasma'' negative mass conventions lead to Vlasov--Poisson systems that are mathematically equivalent to \eqref{VP} under the present Bondi and gravitational-charge conventions, respectively. 
The works \cite{Manfredi2018,Manfredi2026} also explore spatially homogeneous steady states $\rho_+=\rho_-$ and the linear Vlasov--Poisson response around them. These steady states differ from the ones studied here, which are characterized by spatially inhomogeneous finite-mass distributions with compact spatial support and a nontrivial self-consistent radial potential.

We now record the steady-state equations for the Vlasov--Poisson system~\eqref{VP}. 

\begin{mydef}[Steady Vlasov--Poisson system] 
For the Vlasov--Poisson system~\eqref{VP}, a steady state is a time-independent triple
\(\lr{f_+,f_-,\Phi}\) satisfying
\sys{
&\bol{v}\cdot\frac{\p f_+}{\p\bol{x}}
-\frac{m_+^P}{m_+^I}\nabla\Phi\cdot\frac{\p f_+}{\p\bol{v}}=0,\label{fps}\\
&\bol{v}\cdot\frac{\p f_-}{\p\bol{x}}
-\frac{m_-^P}{m_-^I}\nabla\Phi\cdot\frac{\p f_-}{\p\bol{v}}=0,\label{fns}\\
&\Delta\Phi=4\pi G\lr{m_+^A\rho_++m_-^A\rho_-}.\label{Peqs}
}{VPsteady}
\end{mydef}



\subsection{Statement of the main result}
\label{subsec:main}

In this work, we construct spherically symmetric, compactly supported steady states of system \eqref{VP}
from isotropic energy-cutoff distribution functions, defined as follows:  

\begin{mydef}[Energy-cutoff distribution function]
Let
\eq{
H_{\pm}\lr{\bol{x},\bol{v}}
=
\frac{1}{2}\abs{\bol{v}}^2
+
\frac{m_{\pm}^P}{m_{\pm}^I}\Phi
}
denote the particle energy per unit inertial mass of each species. We consider steady  energy-cutoff distribution functions of the form
\eq{
f_{\pm}^{e}\lr{\bol{x},\bol{v}}
=
A_{\pm}
\lr{E_{\pm}-H_{\pm}}^{n_{\pm}}_{\geq 0},\label{ecd}
}
where $A_{\pm}>0$ and $E_{\pm}$ are real constants, and the real cutoff exponents satisfy $n_{\pm}>-1$. Here, the truncated power is defined by
\begin{equation}
\lr{g}_{\geq0}^{n}
=
\begin{cases}
g^n, & g>0,\\
0,   & g\leq0.
\end{cases}
\end{equation}
\end{mydef} 

The first main result concerns 
the Bondi sign convention:

\begin{thm}[Compact Bondi--Vlasov--Poisson equilibria]
\label{thm:BondiCompact}
The steady Vlasov--Poisson system~\eqref{VPsteady} under the Bondi sign convention
$m_{\pm}^I=m_{\pm}^P=m_{\pm}^A=\pm m$
admits nontrivial spherically symmetric, compactly supported steady states.
More precisely, there exist radial potentials $\Phi\lr{r}$, with
$r=\abs{\bol{x}}$, and energy-cutoff distribution functions \eqref{ecd}
such that the triple $\lr{f_+^e,f_-^e,\Phi}$ solves system~\eqref{VPsteady}
in $\mathbb{R}^3$.
For the cutoff exponents $n_+=n_-=-1/2$, the constructed solution satisfies
$\Phi\in C^1\lr{\mathbb{R}^3}$ and
$\rho_+,\rho_-\in C\lr{\mathbb{R}^3}$.
For the cutoff exponents $n_+=n_-=1/2$, the constructed solution satisfies
$\Phi,\rho_+,\rho_-\in C^1\lr{\mathbb{R}^3}$.
\end{thm}

\begin{remark}[Structure of Bondi--Vlasov--Poisson equilibria] 
The steady equilibria obtained in theorem~\ref{thm:BondiCompact} have the following spatial structure:
\[
\text{overlap core}
\quad+\quad
\text{positive shell}
\quad+\quad
\text{exterior vacuum}.
\] 
Specifically, there exist radii \(0<r_1<R\) such that both species have positive
density for \(0\leq r<r_1\), only the positive species has positive density
for \(r_1\leq r<R\), and both species vanish for \(r\geq R\). Accordingly,
the spatial supports of the negative and positive species are
\(0\leq r\leq r_1\) and \(0\leq r\leq R\), respectively. 
\end{remark}

The second main result concerns the gravitational-charge sign convention:

\begin{thm}[Compact gravitational-charge--Vlasov--Poisson equilibria]
\label{thm:GCCompact}
The steady Vlasov--Poisson system~\eqref{VPsteady} under the gravitational-charge sign convention $m_{\pm}^I=m_+^P=m_+^A=-m_-^P=-m_-^A=m$ 
admits nontrivial spherically symmetric, compactly supported steady states.
More precisely, there exist radial potentials $\Phi\lr{r}$, with
$r=\abs{\bol{x}}$, and energy-cutoff distribution functions \eqref{ecd}
such that the triple $\lr{f_+^e,f_-^e,\Phi}$ solves system~\eqref{VPsteady}
in $\mathbb{R}^3$.
For the cutoff exponents $n_+=n_-=-1/2$, the constructed solution satisfies
$\Phi\in C^1\lr{\mathbb{R}^3}$ and
$\rho_+,\rho_-\in C\lr{\mathbb{R}^3}$.
For the cutoff exponents $n_+=n_-=1/2$, the constructed solution satisfies
$\Phi,\rho_+,\rho_-\in C^1\lr{\mathbb{R}^3}$.
\end{thm}

\begin{remark}[Structure of gravitational-charge--Vlasov--Poisson equilibria] 
The steady equilibria obtained in theorem~\ref{thm:GCCompact} have the following spatial structure: 
\[
\text{negative core}
\quad+\quad
\text{vacuum gap}
\quad+\quad
\text{positive shell}
\quad+\quad
\text{exterior vacuum}.
\]
That is, there exist radii \(0<r_1<r_2<R\) such that only the negative
species has positive density for \(0\leq r<r_1\), both species vanish for
\(r_1\leq r\leq r_2\), only the positive species has positive density for
\(r_2<r<R\), and both species vanish for \(r\geq R\). Accordingly, the
spatial supports of the negative and positive species are
\(0\leq r\leq r_1\) and \(r_2\leq r\leq R\), respectively. 
\end{remark}

The proofs of theorems~\ref{thm:BondiCompact} 
and~\ref{thm:GCCompact}
are given in Sec.~\ref{sec:compacteq}, where we construct the equilibria  explicitly. 


\subsection{Interpretation of the main result and astrophysical implications}

Before presenting the details of the proofs of the main theorems, we explain the principal results of this work and briefly discuss their astrophysical and cosmological implications. 

In the 
Bondi model, we find that a pure negative mass gas cannot form a nontrivial 
compact equilibrium: its self-interaction is repulsive, and a virial-type
argument excludes a stationary localized state with finite mass and
second moment. Mixed equilibria nevertheless exist when a suitable
positive mass component is present. Their support consists of an overlap
core, in which both species are present, surrounded by a positive mass
shell and an exterior vacuum region. Thus, although Bondi negative mass
cannot confine itself, it can be spatially confined within a
self-consistent mixed equilibrium.

For the cutoff exponent $n=-1/2$, the density depends linearly on the
potential, and the reduced radial Poisson equation is piecewise
linear. The overlap core, the positive mass shell, and the exterior
vacuum solution can then be matched explicitly. For the regular cutoff $n=1/2$, the distribution functions are bounded
and vanish at the cutoff, and the matter regions are governed by
Lane--Emden-type equations, as in the classical theory of stellar
polytropes \cite{TaruyaSakagami2005}. The same overlap-core/positive shell structure is obtained
through a corresponding radial matching problem.

The gravitational-charge model leads to a qualitatively different
spatial organization. In this case, compact equilibria with an exterior
vacuum are segregated into a negative mass core, a vacuum gap, and a
positive mass shell. For $n=-1/2$, the complete core--gap--shell solution
can again be obtained explicitly. For $n=1/2$, the construction is
formulated in terms of matched Lane--Emden-type core and shell problems,
together with the finite-gap and exterior vacuum conditions required for
a compact configuration. 

The contrast between the two models shows that
the relation among inertial, passive, and active gravitational mass
determines whether positive and negative masses overlap or segregate in
equilibrium. 
In both models, the exterior Newtonian potential depends only on the net
signed mass, while the internal distribution of the two species depends
on the gravitational sign convention. Astrophysical systems with comparable exterior
fields may therefore possess substantially different internal
structures. Hence, the present results provide a kinetic-theory framework for
investigating how negative mass, if present, could be organized in
stellar-dynamical and cosmological systems.

We also record the formal first and second variations of the
energy--Casimir functionals associated with the constructed equilibria.
The resulting quadratic forms identify the sign structure relevant to a
possible stability analysis. The purpose of the present work, however,
is the existence and spatial classification of compact steady states,
rather than the derivation of a dynamical or nonlinear stability
theorem.

\subsection{Organization of this paper}

The present paper is organized as follows. 
In Sec.~\ref{sec:compacteq}, we construct the spherically symmetric equilibria
explicitly. We first establish the obstruction to
compact pure Bondi negative mass steady states, then construct Bondi
equilibria with an overlap-core/positive shell structure, and finally
construct gravitational-charge equilibria with a
negative core/vacuum gap/positive shell structure. In
Sec.~\ref{sec:EC}, we record the formal first and second variations of
the energy--Casimir functionals for both models, emphasizing the sign
structure of the resulting quadratic forms. Finally, in
Sec.~\ref{sec:oriented-mass}, we discuss a possible geometric
interpretation of the Bondi sign convention.

\subsection{Acknowledgments} 
The research of N.S. was partially supported by JSPS KAKENHI Grant 
No. 25K07267, No. 22H04936, and No. 24K00615. 

Generative artificial intelligence tools were used to assist with language editing and to cross-check some of the calculations presented in this work. All scientific content, derivations, and conclusions were independently verified by the author.

\section{Compact Equilibria}\label{sec:compacteq}
The aim of this section is to construct compact equilibria corresponding to
steady-state solutions of the Vlasov--Poisson system \eqref{VP}, thereby proving Theorems~\ref{thm:BondiCompact} and~\ref{thm:GCCompact}. 

We begin by explaining the behavior of purely negative mass steady
distributions in the Bondi and gravitational-charge settings. 

\subsection{Compact pure negative mass equilibria}

As one may expect from the pairwise repulsive behavior, a pure Bondi negative mass distribution  
does not admit compactly supported steady states. This behavior is explained in the following remark: 

\begin{remark}[{Obstruction to pure Bondi negative mass steady states}]
\label{NoBondi}
Consider a pure Bondi negative mass species. Since $f_+=0$,  
assuming $\lim_{\abs{\bol{x}}\rightarrow+\infty}\Phi=0$, the solution of the Poisson equation \eqref{VPP} is
\eq{
\Phi\lr{\bol{x},t}
=
Gm\int_{\mathbb R^3}
\frac{\rho_-\lr{\bol{x}',t}}
{\abs{\bol{x}-\bol{x}'}}\,d\bol{x}'
\geq 0.
}
Hence, the self-interaction is repulsive,  resulting in an obstruction to steady states. To see this, suppose that the distribution has finite spatial second moment 
\eq{
I_-(t)
=
\int_{\mathbb R^6}
\abs{\bol{x}}^2
f_-\lr{\bol{x},\bol{v},t}\,
d\bol{x}d\bol{v}<+\infty.
}
Assuming sufficient regularity and decay at infinity, integrations by parts gives 
\eq{
\frac{d^2 I_-}{dt^2}
=
2\int_{\mathbb R^6}
\abs{\bol{v}}^2 f_-\,d\bol{x}d\bol{v}
-
2\int_{\mathbb R^3}
\rho_-\,\bol{x}\cdot\nabla\Phi\,d\bol{x}.
}
Using the expression for $\Phi$ and symmetrizing in $\bol{x}$ and $\bol{x}'$, 
\eq{
-2\int_{\mathbb R^3}
\rho_-\,\bol{x}\cdot\nabla\Phi\,d\bol{x}
=&
-Gm\int_{\mathbb R^6}
\rho_-\rho_-'
\lrs{
\bol{x}\cdot\nabla
\lr{\frac{1}{\abs{\bol{x}-\bol{x}'}}}
+
\bol{x}'\cdot\nabla'
\lr{\frac{1}{\abs{\bol{x}-\bol{x}'}}}
}
\,d\bol{x}d\bol{x}'
\\
=&
-Gm\int_{\mathbb R^6}
\rho_-\rho_-'
\nabla
\lr{\frac{1}{\abs{\bol{x}-\bol{x}'}}}
\cdot
\lr{\bol{x}-\bol{x}'}
\,d\bol{x}d\bol{x}'
\\
=&
Gm\int_{\mathbb R^6}
\frac{\rho_-\rho_-'}
{\abs{\bol{x}-\bol{x}'}}
\,d\bol{x}d\bol{x}',
}
where $\rho_-'=\rho_-\lr{\bol{x}',t}$ and
$\nabla'=\p/\p\bol{x}'$. It follows that
\eq{
\frac{d^2 I_-}{dt^2}
=
2\int_{\mathbb R^6}
\abs{\bol{v}}^2 f_-\,d\bol{x}d\bol{v}
+
Gm\int_{\mathbb R^6}
\frac{\rho_-\rho_-'}
{\abs{\bol{x}-\bol{x}'}}
\,d\bol{x}d\bol{x}'
>0,
\label{It2g0}
}
for every nontrivial distribution with finite kinetic and interaction energies. 
At equilibrium 
$d^2I_-/dt^2=0$, contradicting \eqref{It2g0}. Thus, pure Bondi negative mass steady states with finite spatial second moment and finite kinetic and interaction energies cannot exist. 
In particular, a nontrivial  compactly supported steady state made of pure Bondi negative mass is impossible.
\end{remark}

\begin{remark}[{Pure gravitational-charge negative mass steady states}]
Consider a pure gravitational-charge negative mass species, and define 
$\Psi=-\Phi$. Then, the governing Vlasov--Poisson system can be written as  
\eq{
\frac{\p f_-}{\p t}
+\bol{v}\cdot\frac{\p f_-}{\p\bol{x}}
-\nabla\Psi\cdot\frac{\p f_-}{\p\bol{v}}
=0,
\qquad
\Delta\Psi=4\pi Gm\rho_- .
}
It follows that a pure gravitational-charge negative mass species is mathematically equivalent
to the standard attractive gravitational Vlasov--Poisson system. In particular, standard compact steady states, such
as isotropic energy-cutoff equilibria, can be constructed in the same way as
for ordinary positive mass.

\end{remark}

\subsection{Density formula, Poisson equation, and exterior vacuum potential}

Our next goal is to construct compactly supported steady states of system \eqref{VP} involving both positive and negative mass. To this end, we first derive the spatial densities corresponding to the energy-cutoff distributions \eqref{ecd}. Setting $v=\abs{\bol{v}}$ and $\chi_{\pm}=m_{\pm}^P/m_{\pm}^I$, we obtain
\eq{
\rho_{\pm}
&=4\pi A_{\pm}\int_0^{\sqrt{2\lr{E_{\pm}-\chi_{\pm}\Phi}_{\geq 0}}}
\lr{E_{\pm}-\frac{v^2}{2}-\chi_{\pm}\Phi}^{n_{\pm}}v^2\,dv\\
&=4\pi A_{\pm}2^{3/2}
\lr{E_{\pm}-\chi_{\pm}\Phi}_{\geq 0}^{n_{\pm}+3/2}
\int_0^1\lr{1-y^2}^{n_{\pm}}y^2\,dy.
}
The remaining integral is a beta integral. Setting \(s=y^2\), we have
\(dy=\frac12s^{-1/2}ds\), and hence
\eq{
\int_0^1\lr{1-y^2}^{n_{\pm}}y^2\,dy
=
\frac12\int_0^1\lr{1-s}^{n_{\pm}}s^{1/2}\,ds
=
\frac12 B\lr{\frac32,n_{\pm}+1}.
}
Using
\(B(a,b)=\frac{\Gamma(a)\Gamma(b)}{\Gamma(a+b)}\)
and \(\Gamma(3/2)=\sqrt{\pi}/2\), we find
\eq{
\int_0^1\lr{1-y^2}^{n_{\pm}}y^2\,dy
=
\frac{\sqrt{\pi}}{4}
\frac{\Gamma(n_{\pm}+1)}{\Gamma(n_{\pm}+5/2)}.
}
Therefore,
\eq{
\rho_{\pm}
&=
C_{\pm}\lr{E_{\pm}-\chi_{\pm}\Phi}^{n_{\pm}+3/2}_{\geq0},
\qquad
C_{\pm}
=
2^{3/2}\pi^{3/2}A_{\pm}
\frac{\Gamma(n_{\pm}+1)}{\Gamma(n_{\pm}+5/2)},
\label{rhopm}
}
where \(\Gamma\) denotes the gamma function. The static Poisson equation \eqref{Peqs} can thus be written as
\eq{
\Delta\Phi
=
4\pi G m\lrs{
C_+\lr{E_+-\chi_+\Phi}_{\geq 0}^{n_++3/2}
-
C_-\lr{E_--\chi_-\Phi}_{\geq 0}^{n_-+3/2}
}.
\label{Peq2}
}

In the following, we construct spherically symmetric solutions of system \eqref{VPsteady} for which matter is confined within a finite radius $r=\abs{\bol{x}}\leq R$. For all the solutions considered here, the outer interface $r=R$ is characterized by
\(\Phi(R)=E_+\), so that \(\rho_+(R)=0\). Since no matter is present for \(r>R\), the potential \(\Phi\lr{r}\) is harmonic in the exterior vacuum region. Hence,
\eq{
\Phi(r)
=
E_+
+
Q\left(\frac1R-\frac1r\right),
\qquad
r\ge R,
\label{Phiext}
}
where the constant $Q$ is obtained by integrating the Poisson equation over a sphere of radius $R$:
\eq{
Q=R^2\Phi'(R)
=
G(M_+-M_-),\label{QdM}
}
with
\eq{
M_{\pm}
=
\abs{m_{\pm}^A}\int_{\mathbb{R}^3}\rho_{\pm}\,d\bol{x}
=
m\int_{\mathbb{R}^3}\rho_{\pm}\,d\bol{x},
}
the absolute values of the total active masses.


\subsection{Compact Bondi--Vlasov--Poisson equilibria}\label{subsec:BVPeq}
We seek for compactly supported 
solutions $f_{\pm}\lr{\bol{x},\bol{v}}$ with self-consistent gravitational potential $\Phi\lr{\bol{x}}$ of the steady Vlasov--Poisson system 
\eqref{VPsteady} with Bondi negative mass. 
Assume that $E_+>E_-$. 
From equation \eqref{rhopm}, we see that the support structure of the solution is
\sys{
&\Phi<E_-:
\qquad
\rho_+>0,\quad \rho_->0,\\
&
E_-\le \Phi< E_+:
\qquad
\rho_+>0,\quad \rho_-=0,\\
&\Phi\ge E_+:
\qquad
\rho_+=\rho_-=0.
}{support}
We look for a spherically symmetric potential $\Phi\lr{r}$ with interfaces $0<r_1<R$ 
such that
\sys{
&0\le r<r_1:
\quad
\Phi(r)<E_-\qquad \text{(overlap core)},\\
&r_1<r<R:
\quad
E_-<\Phi(r)<E_+\qquad\text{(positive shell)},\\
&r>R:
\quad
\Phi(r)\ge E_+\qquad\text{(exterior vacuum)}.
}{interfaces}
The interfaces are determined by the level sets
$\Phi\lr{r_1}=E_-$ and $\Phi\lr{R}=E_+$.  
Both the potential and its normal derivative must be continuous across each interface, $
[\Phi]=\left[\Phi'\right]=0$, 
where $\Phi'={d\Phi}/{dr}$. 
These conditions determine the integration constants together with the regularity condition at the origin 
$\Phi'(0)=0$.

\subsubsection{The case $n=-1/2$}

Set $
n_+=n_-=n=-1/2$. Then, the constants in the density formula \eqref{rhopm} are $C_\pm=2^{3/2}\pi^2 A_\pm$. 
We define 
\eq{
a=4\pi Gm C_+
=
2^{7/2}\pi^3Gm A_+,
\qquad
b=4\pi Gm C_-
=
2^{7/2}\pi^3Gm A_-.
}
With this notation, the Poisson equation \eqref{Peq2} takes the linear form
\eq{
\Delta\Phi
=
a\lr{E_+-\Phi}_{\geq 0}
-
b\lr{E_--\Phi}_{\geq 0}.\label{Peq3}
}
We set \(\delta=E_+-E_->0\), 
\(\eta=E_- -\Phi_c\), 
and $\Phi_c=\Phi\lr{0}$, and solve the Poisson equation \eqref{Peq3} in each region.  


\paragraph{Overlap core.}

For \(0\leq r<r_1\) we have \(\Phi<E_-\). Hence, both species are present. 
To keep the construction explicit, we restrict to the case $a=b$, 
which is equivalent to choosing \(A_+=A_-\). Then \eqref{Peq3} reduces to
\eq{
\Delta\Phi=a\lr{E_+-E_-}=a\delta.
}
We impose regularity at the origin by requiring
$\Phi(0)=\Phi_c<E_-$ and $\Phi'(0)=0$. 
The resulting regular core solution is
\eq{
\Phi(r)
=
\Phi_c+\frac{a\delta}{6}r^2.
}
The first interface \(r_1\) is defined by 
$\Phi(r_1)=E_-$. Since \(\eta=E_--\Phi_c>0\), we obtain
\eq{
r_1^2
=
\frac{6\eta}{a\delta}.
}
For later convenience, define $p_1=\Phi'(r_1)$. 
We have $p_1={a\delta}r_1/3$, and 
$r_1p_1=2\eta$.

\paragraph{Positive shell.}

For $r_1<r<R$, we have $E_-<\Phi<E_+$. Hence, \(\rho_+>0\) and \(\rho_-=0\). It follows that  
\eq{
\Delta\Phi=a\lr{E_+-\Phi}.
}
Set $y(r)=E_+-\Phi(r)>0$. Then, the Poisson equation reduces to 
\eq{
y''+\frac{2}{r}y'+ay=0.\label{yr}
}
The matching conditions at \(r=r_1\) are
\eq{
y(r_1)=E_+-E_-=\delta,
\qquad
y'(r_1)=-\Phi'(r_1)=-p_1.
}
The positive shell potential for \(r_1\leq r\leq R\) thus takes the form 
\eq{
\Phi(r)
=
E_+-\frac{r_1\delta}{r}
\cos\lrs{\sqrt{a}\lr{r-r_1}}
-\frac{\delta-r_1p_1}{r\sqrt{a}}
\sin\lrs{\sqrt{a}\lr{r-r_1}}.
}
We must now show that the outer interface $R$ exists. To this end, we change variables again to  
$z(r)=ry(r)$ so that eq.~\eqref{yr} and the matching conditions at $r=r_1$ become  
\eq{
z''+az=0,
\qquad
z(r_1)=r_1\delta,
\qquad
z'(r_1)=\delta-r_1p_1.
}
If $r_1p_1\geq\delta$,
then \(z'(r_1)\leq0\). Since \(z''=-az<0\) while \(z>0\), the function \(z'\) is strictly decreasing, and \(z\) reaches zero at a finite first radius \(R>r_1\). Since $r_1p_1=2\eta$, 
the sufficient condition $r_1 p_1\geq \delta$ is equivalent to $\eta\geq {\delta}/{2}$. 
Since \(z\) decreases from \(z(r_1)=r_1\delta\) until its first zero
\(R\), we have \(0<y(r)=z(r)/r<\delta\) for \(r_1<r<R\).
Hence \(E_-<\Phi(r)<E_+\), so the positive shell equation is
self-consistent. 
The outer interface is the first radius \(R>r_1\) such that $y(R)=0$, 
or equivalently \(\Phi(R)=E_+\). Recalling eq.~\eqref{QdM},   
we find \(Q=-Rz'(R)\). At the first zero of \(z\),
\(z'(R)<0\), and therefore $Q>0$. 
Hence, the exterior solution \eqref{Phiext} satisfies \(\Phi(r)\geq E_+\) for \(r\geq R\), and the exterior region remains vacuum.

\paragraph{Complete solution.}

\begin{figure}[h!]
    \centering
    \includegraphics[width=0.65\textwidth]{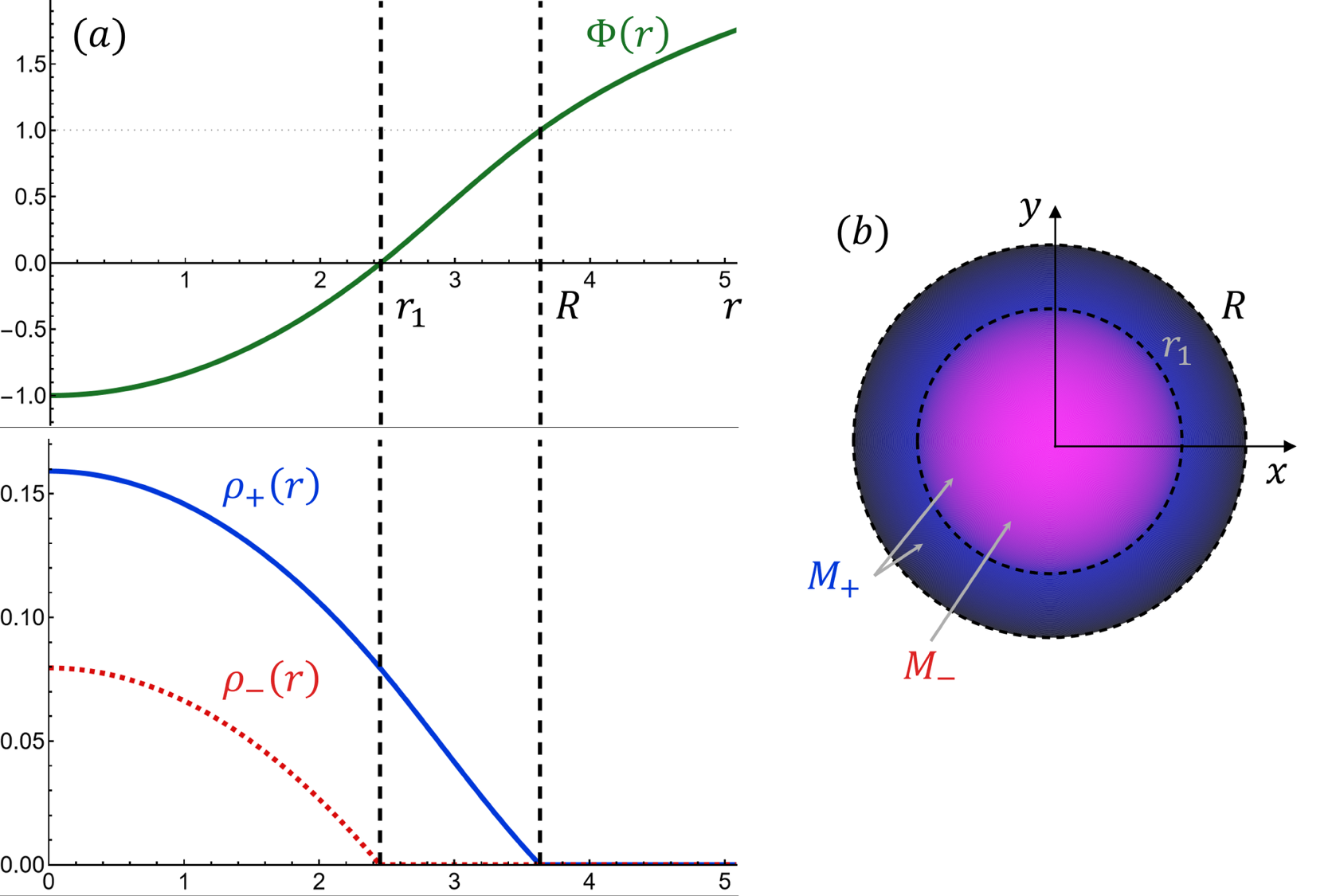}
    \caption{
    Compact Bondi--Vlasov--Poisson equilibrium for the exactly solvable
    cutoff exponent \(n_+=n_-=-1/2\). The plotting parameters are
    \(G=m=1\), \(E_+=1\), \(E_-=0\), \(\Phi_c=-1\), and \(a=b=1\). Thus
    \(\delta=E_+-E_-=1\) and \(C_+=C_-=1/4\pi\). The corresponding interfaces are
    \(r_1\simeq2.449\) and \(R\simeq3.633\).
    (a) The upper panel shows the matched potential \(\Phi(r)\), while the
    lower panel shows the density profiles
    \(\rho_+(r)=C_+(E_+-\Phi(r))_{\geq0}\) and
    \(\rho_-(r)=C_-(E_--\Phi(r))_{\geq0}\). The blue solid curve denotes
    \(\rho_+\), and the red dotted curve denotes \(\rho_-\). The vertical
    dashed lines mark the interfaces \(r=r_1\) and \(r=R\).
    (b) A two-dimensional \(x\)-\(y\) cross-section of the density
    distribution. Blue represents the positive mass \(M_+\), red
    represents the negative mass \(M_-\), and purple indicates the
    overlap region where both species are present. The color intensities in
    panel (b) are normalized separately by the maximum of each species density
    for visualization.
    }
    \label{fig1}
\end{figure}

Fix \(E_+>E_-\), set $\delta=E_+-E_->0$, 
choose \(\Phi_c<E_-\), and let 
$\eta=E_--\Phi_c\geq{\delta}/{2}$.
For \(a=b\), define $r_1=\sqrt{{6\eta}/{a\delta}}$ and 
$p_1={a\delta}r_1/3$. 
Let \(R>r_1\) be the first zero of \(y=E_+-\Phi\), and define $Q=R^2\Phi'(R)=-R^2y'(R)>0$. 
Then the full matched potential is
\begin{equation}\label{eq:Phi-piecewise-nminus-half}
\Phi(r)
=
\begin{cases}
\displaystyle
\Phi_c+\frac{a\delta}{6}r^2,
& 0\leq r\leq r_1,
\\[2.0ex]
\displaystyle
E_+
-
\frac{r_1\delta}{r}
\cos\lrs{\sqrt{a}\lr{r-r_1}}
-
\frac{\delta-r_1p_1}{r\sqrt{a}}
\sin\lrs{\sqrt{a}\lr{r-r_1}},
& r_1\leq r\leq R,
\\[2.0ex]
\displaystyle
E_+
+
Q\left(\frac1R-\frac1r\right),
& r\geq R.
\end{cases}
\end{equation}
With these definitions, \(\Phi\) and \(\Phi'\) are continuous at \(r=r_1\) and \(r=R\). Moreover,
$Q=R^2\Phi'(R)=G\lr{M_+-M_-}>0$. 
The active masses \(M_{\pm}\) are finite because the solution is compactly supported. A plot of the solution \eqref{eq:Phi-piecewise-nminus-half} for a specific set of parameters is shown in fig.~\ref{fig1}.

\begin{remark}
The restriction \(a=b\) is made to keep the construction and matching
calculations as simple as possible. The cases \(a\neq b\) can be treated
in a similar fashion, but are not presented here.
\end{remark}



\subsubsection{The case $n=1/2$}

Set \(n_+=n_-=n=1/2\). Then, the constants in the density formula~\eqref{rhopm} are
\(C_\pm={\pi^2}A_\pm/{\sqrt{2}}\). We define
\eq{
a=4\pi Gm C_+
=
2\sqrt{2}\pi^3Gm A_+,
\qquad
b=4\pi Gm C_-
=
2\sqrt{2}\pi^3Gm A_-.
}
With this notation, the Poisson equation \eqref{Peq2} becomes
\eq{
\Delta\Phi
=
a\lr{E_+-\Phi}_{\geq0}^2
-
b\lr{E_--\Phi}_{\geq0}^2 .
}
We set \(\delta=E_+-E_->0\), \(\eta=E_--\Phi_c>0\), and
\(\Phi_c=\Phi(0)<E_-\), and solve the Poisson equation in each region.

\paragraph{Overlap core.}

For \(0\leq r<r_1\), we have \(\Phi<E_-\). Hence, both species are present.
To keep the construction explicit, we restrict to the case \(a=b\), which is
equivalent to choosing \(A_+=A_-\). 
Setting
\(\Phi_\ast=\lr{E_++E_-}/2\) and \(\lambda^2=2a\delta\), the core equation becomes
\eq{
\Delta\lr{\Phi-\Phi_\ast}
+
\lambda^2\lr{\Phi-\Phi_\ast}
=0.
}
We impose regularity at the origin by requiring
\(\Phi(0)=\Phi_c<E_-\) and \(\Phi'(0)=0\). The resulting regular core solution is
\eq{
\Phi(r)
=
\Phi_\ast
+
\lr{\Phi_c-\Phi_\ast}
\frac{\sin\lr{\lambda r}}{\lambda r}.
}
The first interface \(r_1\) is defined by \(\Phi(r_1)=E_-\). Setting
\(x_1=\lambda r_1\), this condition gives
\eq{
\frac{\sin x_1}{x_1}
=
\frac{E_--\Phi_\ast}{\Phi_c-\Phi_\ast}
=
\frac{\delta}{2\eta+\delta}.
}
Since \(0<\delta/\lr{2\eta+\delta}<1\), while \(\sin x/x\) decreases
continuously from \(1\) to \(0\) on \(0<x<\pi\), there exists a unique
\(x_1\in(0,\pi)\). Hence \(r_1=x_1/\lambda\) is well-defined. 
For later convenience, define \(p_1=\Phi'(r_1)\). Differentiating the core
solution gives
\eq{
r_1p_1
=
\left(\eta+\frac{\delta}{2}\right)
\frac{\sin x_1-x_1\cos x_1}{x_1}.
}
As \(\eta\rightarrow+\infty\), the interface equation implies
\(x_1\rightarrow\pi^-\), while
\(\lr{\sin x_1-x_1\cos x_1}/x_1\rightarrow1\). Therefore
\(r_1p_1\rightarrow+\infty\), and the central depth \(\eta\) can be chosen
so that
\eq{
r_1p_1\geq\delta.
}

\paragraph{Positive shell.}

For \(r_1<r<R\), we have \(E_-<\Phi<E_+\). Hence,
\(\rho_+>0\) and \(\rho_-=0\), and
\eq{
\Delta\Phi=a\lr{E_+-\Phi}^2.
}
Set \(y(r)=E_+-\Phi(r)>0\). Then, the Poisson equation reduces to the
Lane--Emden-type equation
\eq{
y''+\frac{2}{r}y'+ay^2=0.
}
The matching conditions at \(r=r_1\) are
\(y(r_1)=\delta\) and \(y'(r_1)=-p_1\). 
We must now show that the outer interface \(R\) exists. To this end, set
\(z(r)=ry(r)\). Then
\eq{
z''=-\frac{a}{r}z^2,
\qquad
z(r_1)=r_1\delta,
\qquad
z'(r_1)=\delta-r_1p_1.
}
If \(r_1p_1\geq\delta\), then \(z'(r_1)\leq0\). Since \(z''<0\) while
\(z>0\), the function \(z'\) is strictly decreasing. Hence, \(z\) reaches
zero at a finite first radius \(R>r_1\). Since \(z\) decreases from \(z(r_1)=r_1\delta\) until its first zero
\(R\), we have \(0<y(r)=z(r)/r<\delta\) for \(r_1<r<R\).
Hence \(E_-<\Phi(r)<E_+\), so the positive shell equation is
self-consistent.  
The outer interface is the first radius \(R>r_1\) such that \(y(R)=0\),
or equivalently \(\Phi(R)=E_+\). Recalling \eqref{Phiext}, we have
\(Q=-Rz'(R)\). At the first zero of \(z\), \(z'(R)<0\), and therefore
\(Q>0\). Hence, the exterior solution \eqref{Phiext} satisfies
\(\Phi(r)\geq E_+\) for \(r\geq R\), and the exterior region remains vacuum.

\paragraph{Complete solution.}

\begin{figure}[h!]
    \centering
    \includegraphics[width=0.65\textwidth]{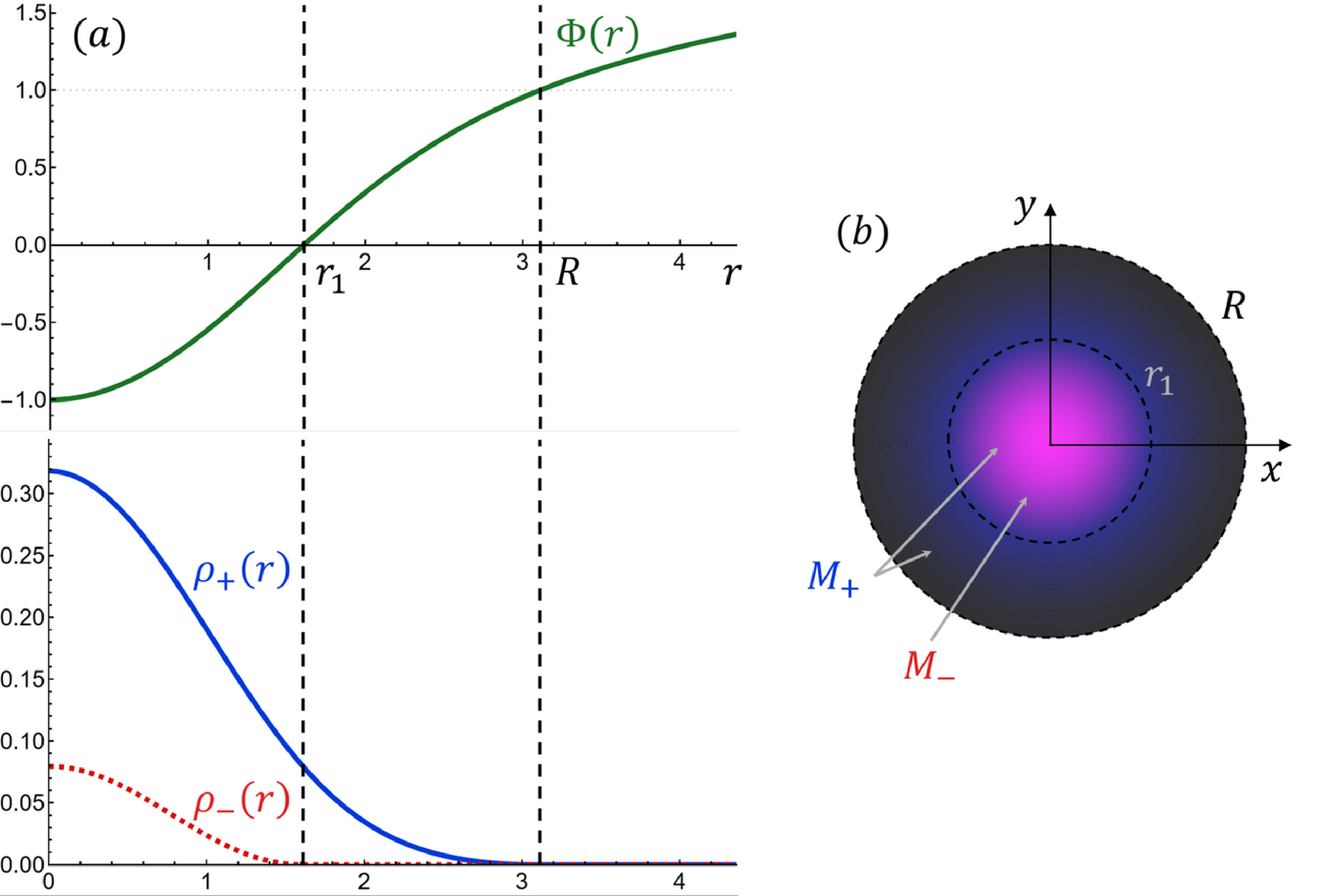}
    \caption{
    Compact Bondi--Vlasov--Poisson equilibrium for the regular cutoff
    exponent \(n_+=n_-=1/2\). The plotting parameters are
    \(G=m=1\), \(E_+=1\), \(E_-=0\), \(\Phi_c=-1\), and \(a=b=1\). Thus
    \(\delta=E_+-E_-=1\), \(\eta=E_- - \Phi_c=1\),
    \(\Phi_\ast=(E_++E_-)/2=1/2\), \(\lambda=\sqrt{2}\), and
    \(C_+=C_-=1/4\pi\). The corresponding interfaces are
    \(r_1\simeq1.611\) and \(R\simeq3.114\).
    (a) The upper panel shows the matched potential \(\Phi(r)\), while the
    lower panel shows the density profiles
    \(\rho_+(r)=C_+(E_+-\Phi(r))_{\geq0}^{2}\) and
    \(\rho_-(r)=C_-(E_--\Phi(r))_{\geq0}^{2}\). The blue solid curve denotes
    \(\rho_+\), and the red dotted curve denotes \(\rho_-\). The vertical
    dashed lines mark the interfaces \(r=r_1\) and \(r=R\).
    (b) A two-dimensional \(x\)-\(y\) cross-section of the density
    distribution. Blue represents the positive mass \(M_+\), red
    represents the negative mass \(M_-\), and purple indicates the
    overlap region where both species are present. The color intensities in
    panel (b) are normalized separately by the maximum of each species density
    for visualization.
    }
    \label{fig2}
\end{figure}

Fix \(E_+>E_-\), set \(\delta=E_+-E_->0\), and choose
\(E_->\Phi_c\), or equivalently \(\eta=E_--\Phi_c>0\), sufficiently large
so that \(r_1p_1\geq\delta\). For \(a=b\), set
\(\Phi_\ast=\lr{E_++E_-}/2\), \(\lambda^2=2a\delta\), let
\(x_1\in(0,\pi)\) be the unique solution of
\(\sin x_1/x_1=\delta/\lr{2\eta+\delta}\), and define
\(r_1=x_1/\lambda\). The slope at the first interface is
\[
p_1
=
\lambda\lr{\Phi_c-\Phi_\ast}
\frac{x_1\cos x_1-\sin x_1}{x_1^2}.
\]
Let \(y\) be the positive shell solution of the Lane--Emden-type equation $y''+2y'/r+ay^2=0$ with $y(r_1)=\delta$ and $y'\lr{r_1}=-p_1$, \(R>r_1\) its first zero,
and define \(Q=-R^2y'(R)>0\). Then the full matched potential is
\begin{equation}\label{eq:Phi-piecewise-nhalf}
\Phi(r)
=
\begin{cases}
\displaystyle
\Phi_\ast
+
\lr{\Phi_c-\Phi_\ast}
\frac{\sin(\lambda r)}{\lambda r},
& 0\le r\le r_1,
\\[2.0ex]
\displaystyle
E_+-y(r),
& r_1\le r\le R,
\\[2.0ex]
\displaystyle
E_+
+
Q\left(\frac1R-\frac1r\right),
& r\ge R.
\end{cases}
\end{equation}
With these definitions, \(\Phi\) and \(\Phi'\) are continuous at
\(r=r_1\) and \(r=R\). Moreover,
\(Q=R^2\Phi'(R)=G\lr{M_+-M_-}>0\), and the active masses \(M_\pm\) are
finite because the solution is compactly supported. A plot of the solution
\eqref{eq:Phi-piecewise-nhalf} for a specific set of parameters is shown in
fig.~\ref{fig2}.

\begin{remark}
The restriction \(a=b\) is made to keep the construction and matching
calculations as simple as possible. The case \(a\neq b\) can be treated
in a similar fashion as a shooting problem, but is not presented here.
\end{remark}

\subsection{Compact gravitational-charge--Vlasov--Poisson equilibria}

We seek for compactly supported 
solutions \(f_{\pm}\lr{\bol{x},\bol{v}}\) with self-consistent gravitational potential \(\Phi\lr{\bol{x}}\) of the steady Vlasov--Poisson system \eqref{VPsteady} with gravitational-charge negative mass. 
It is useful to introduce \(L=-E_-\). 
Assume \(E_+<L\), or equivalently \(E_++E_-<0\), and set
\(\delta=L-E_+>0\). The interval \(E_+\leq\Phi\leq L\) is then a vacuum
window, and the support structure is
\sys{
&\Phi>L:
\qquad
\rho_+=0,\quad \rho_->0,\\
&E_+\leq\Phi\leq L:
\qquad
\rho_+=\rho_-=0,\\
&\Phi<E_+:
\qquad
\rho_+>0,\quad \rho_-=0.
}{supportG} 
We look for a spherically symmetric potential \(\Phi\lr{r}\), and interfaces \(0<r_1<r_2<R\) such that
\sys{
&0\leq r<r_1:
\quad
\Phi(r)>L
\qquad\text{(negative core)},\\
&r_1<r<r_2:
\quad
E_+<\Phi(r)<L
\qquad\text{(vacuum gap)},\\
&r_2<r<R:
\quad
\Phi(r)<E_+
\qquad\text{(positive shell)},\\
&r>R:
\quad
E_+\leq\Phi(r)\leq L
\qquad\text{(exterior vacuum)}.
}{interfacesG}
The interfaces are determined by
\(\Phi(r_1)=L\) and \(\Phi(r_2)=\Phi(R)=E_+\).
Both the potential and its normal derivative must be continuous across each
interface, \([\Phi]=[\Phi']=0\), and satisfy the
regularity condition \(\Phi'(0)=0\).

\subsubsection{The case $n=-1/2$}

Set \(n_+=n_-=n=-1/2\). Then, the constants in the density formula
\eqref{rhopm} are \(C_\pm=2^{3/2}\pi^2A_\pm\). We define
\eq{
a=4\pi Gm C_+
=
2^{7/2}\pi^3Gm A_+,
\qquad
b=4\pi Gm C_-
=
2^{7/2}\pi^3Gm A_-.
}
With this notation, the Poisson equation \eqref{Peq2} takes the linear form
\eq{
\Delta\Phi
=
a\lr{E_+-\Phi}_{\geq0}
-
b\lr{\Phi-L}_{\geq0}.
}
To keep the construction explicit, we restrict to the case \(a=b\), which is
equivalent to choosing \(A_+=A_-\), and set \(\kappa=\sqrt a\). We solve the
Poisson equation in each region.

\paragraph{Negative core.}

For \(0\leq r<r_1\), we have \(\Phi>L\). Hence, \(\rho_->0\) and
\(\rho_+=0\), and
\eq{
\Delta\lr{\Phi-L}+a\lr{\Phi-L}=0.
}
Choose a central amplitude \(\mathcal A>0\). Imposing regularity at the origin
gives
\eq{
\Phi(r)
=
L+\mathcal A\frac{\sin(\kappa r)}{\kappa r}.
}
The first interface \(r_1\) is defined by \(\Phi(r_1)=L\), and is therefore $r_1={\pi}/{\kappa}$. 
Since \(\sin(\kappa r)>0\) for \(0<r<r_1\), the negative core condition
\(\Phi>L\) is self-consistent. For later convenience, define
\(P=\mathcal A r_1\). Then
\(\Phi'(r_1)=-\mathcal A/r_1=-P/r_1^2\).

\paragraph{Vacuum gap.}

For \(r_1<r<r_2\), both densities vanish, so \(\Delta\Phi=0\). Matching
\(\Phi\) and \(\Phi'\) at \(r=r_1\) gives
\eq{
\Phi(r)
=
L-P\left(\frac1{r_1}-\frac1r\right).
}
The positive shell begins at the first radius \(r_2>r_1\) such that
\(\Phi(r_2)=E_+\). Since \(\delta=L-E_+\), we obtain $r_2={r_1}/\lr{1-\delta/\mathcal A}$. 
Thus, \(r_2\) is finite and satisfies \(r_2>r_1\) provided that $\mathcal A>\delta$. 
Moreover, \(\Phi\) decreases monotonically from \(L\) to \(E_+\) on
\(r_1<r<r_2\), so the vacuum gap condition \(E_+<\Phi<L\) is
self-consistent.

\paragraph{Positive shell.}

For \(r_2<r<R\), we have \(\Phi<E_+\). Hence, \(\rho_+>0\) and
\(\rho_-=0\). Set \(y(r)=E_+-\Phi(r)>0\). Then, the Poisson equation reduces
to
\eq{
y''+\frac{2}{r}y'+ay=0.
}
The matching conditions at \(r=r_2\) are
\(y(r_2)=0\) and \(y'(r_2)=P/r_2^2\). The corresponding solution is
\eq{
y(r)
=
\frac{P}{\kappa r_2}
\frac{\sin\lrs{\kappa\lr{r-r_2}}}{r}.
}
Thus, \(y>0\), or equivalently \(\Phi<E_+\), for
\(r_2<r<r_2+\pi/\kappa\), and the outer interface is $R=r_2+{\pi}/{\kappa}$. 
At \(r=R\), \(y(R)=0\) and hence \(\Phi(R)=E_+\). Recalling
\eqref{Phiext}, the exterior coefficient is
\(Q=-R^2y'(R)=PR/r_2>0\). The exterior vacuum condition
\(Q/R\leq\delta\) becomes
\(P/r_2=\mathcal A-\delta\leq\delta\), or equivalently
\(\mathcal A\leq2\delta\). Together with the condition
\(\mathcal A>\delta\), the compact construction is therefore valid for
\eq{
0<\delta<\mathcal A\leq2\delta.
}

\paragraph{Complete solution.}

\begin{figure}[h!]
    \centering
    \includegraphics[width=0.65\textwidth]{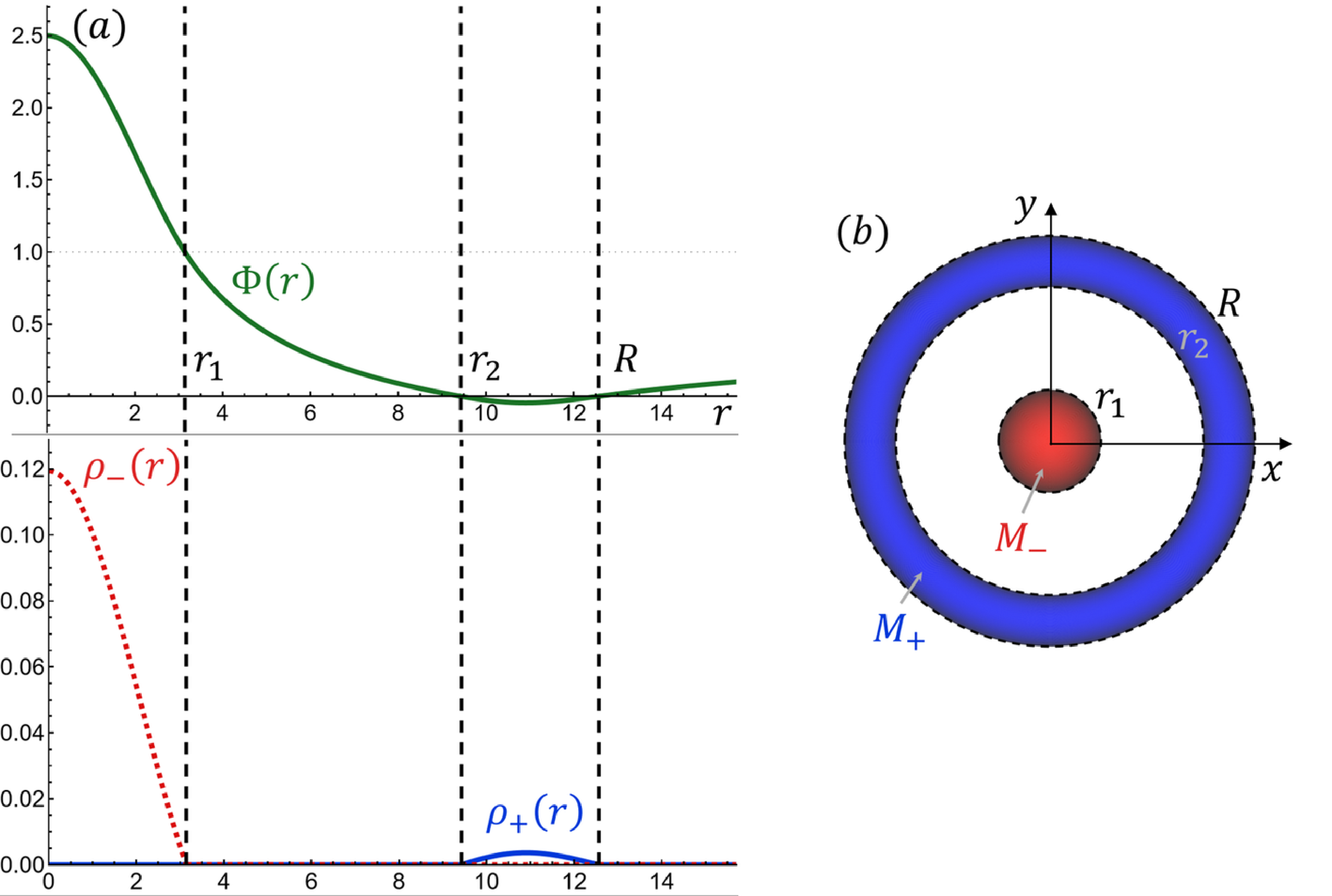}
    \caption{
    Gravitational-charge equilibrium for the compact exactly solvable
    cutoff exponent \(n_+=n_-=-1/2\). The plotting parameters are
    \(G=m=1\), \(a=b=1\), \(E_+=0\), \(E_-=-1\), and hence
    \(L=-E_-=1\) and \(\delta=L-E_+=1\). The core amplitude is
    \(\mathcal A=1.5\). The resulting matching data are
    \(r_1\simeq3.14159\), \(r_2\simeq9.42478\), \(R\simeq12.5664\),
    \(P\simeq4.71239\), and \(Q\simeq6.28319\), with
    \(Q/R=0.5\leq\delta\).
    Panel (a) shows the matched radial potential \(\Phi(r)\) and the density
    profiles
    \(\rho_+(r)=C_+(E_+-\Phi(r))_{\geq0}\) and
    \(\rho_-(r)=C_-(E_-+\Phi(r))_{\geq0}\), with
    \(C_+=C_-=1/4\pi\). The red dotted curve is the negative mass density
    \(\rho_-\), supported in the inner core \(0\leq r\leq r_1\), while the
    blue solid curve is the positive mass density \(\rho_+\), supported in
    the shell \(r_2\leq r\leq R\). The interval \(r_1<r<r_2\) is a vacuum
    gap, and the dashed vertical lines mark the interfaces \(r_1\), \(r_2\),
    and \(R\).
    Panel (b) shows the corresponding two-dimensional \(x\)-\(y\)
    cross-section: the red disk represents the negative mass core \(M_-\),
    the blue annulus represents the positive mass shell \(M_+\), and the
    white region between them is the vacuum gap. The color intensities in
    panel (b) are normalized by the maximum of each species
    density for visualization.
    }
    \label{fig3}
\end{figure}

Choose the core amplitude \(\mathcal A\) so that
\(0<\delta<\mathcal A\leq2\delta\). For \(a=b\), set
\[
\kappa=\sqrt a,\qquad
r_1=\frac{\pi}{\kappa},\qquad
r_2=\frac{r_1}{1-\delta/\mathcal A},\qquad
R=r_2+\frac{\pi}{\kappa},
\qquad
P=\mathcal A r_1,
\qquad
Q=\frac{PR}{r_2}.
\]
Then the full matched potential is
\begin{equation}\label{eq:Phi-piecewise-GVP-nminus-half}
\Phi(r)
=
\begin{cases}
\displaystyle
L+\mathcal A\frac{\sin(\kappa r)}{\kappa r},
& 0\leq r\leq r_1,
\\[2.0ex]
\displaystyle
L-P\left(\frac1{r_1}-\frac1r\right),
& r_1\leq r\leq r_2,
\\[2.0ex]
\displaystyle
E_+
-
\frac{P}{\kappa r_2}
\frac{\sin\lrs{\kappa\lr{r-r_2}}}{r},
& r_2\leq r\leq R,
\\[2.0ex]
\displaystyle
E_+
+
Q\left(\frac1R-\frac1r\right),
& r\geq R.
\end{cases}
\end{equation}
With these definitions, \(\Phi\) and \(\Phi'\) are continuous at
\(r=r_1\), \(r=r_2\), and \(r=R\). Moreover,
\(Q=R^2\Phi'(R)=G\lr{M_+-M_-}>0\), and the active masses \(M_\pm\) are
finite because the solution is compactly supported. A plot of the solution
\eqref{eq:Phi-piecewise-GVP-nminus-half} for a specific set of parameters is
shown in fig.~\ref{fig3}.

\begin{remark}
The restriction \(a=b\) is made to keep the construction and matching
calculations as simple as possible. The case \(a\neq b\) can be treated in a
similar fashion, with distinct core and shell length scales, but is not
presented here.
\end{remark}

\subsubsection{The case $n=1/2$}

Set \(n_+=n_-=n=1/2\). Then, the constants in the density formula
\eqref{rhopm} are \(C_\pm={\pi^2}A_\pm/{\sqrt{2}}\). We define
\eq{
a=4\pi Gm C_+
=
2\sqrt{2}\pi^3Gm A_+,
\qquad
b=4\pi Gm C_-
=
2\sqrt{2}\pi^3Gm A_-.
}
With this notation, the Poisson equation becomes
\eq{
\Delta\Phi
=
a\lr{E_+-\Phi}_{\geq0}^2
-
b\lr{\Phi-L}_{\geq0}^2.
}
To keep the construction explicit, we restrict to the case \(a=b\), which is
equivalent to choosing \(A_+=A_-\). We solve the Poisson equation in each
region.

\paragraph{Negative core.}

For \(0\leq r<r_1\), we have \(\Phi>L\). Hence, \(\rho_->0\) and
\(\rho_+=0\). Set \(\psi(r)=\Phi(r)-L>0\). Then, the Poisson equation
reduces to the Lane--Emden-type equation
\eq{
\psi''+\frac{2}{r}\psi'+a\psi^2=0.
}
Choose a central amplitude \(\Psi_c>0\), with
\(\psi(0)=\Psi_c\) and \(\psi'(0)=0\). Let \(\theta\lr{\xi}\) be the Lane--Emden
profile determined by
\eq{
\theta''+\frac{2}{\xi}\theta'+\theta^2=0,
\qquad
\theta(0)=1,
\qquad
\theta'(0)=0.
}
Then
\eq{
\Phi(r)
=
L+\Psi_c\,\theta\lr{\sqrt{a\Psi_c}\,r}.
}
Let \(\xi_1>0\) be the first zero of \(\theta\), and define
\(\mu=-\xi_1^2\theta'(\xi_1)>0\). 
The first zero exists because, setting \(w=\xi\theta\), one has
\(w''=-w^2/\xi<0\); moreover, as long as \(w>0\), this concavity forces
\(w'\) to become negative at finite \(\xi\), after which \(w\), and hence
\(\theta\), crosses zero at a finite radius. 
The fact that $\theta'\lr{\xi_1}<0$ follows by noting that $\lr{\xi^2\theta'}'=-\xi^2\theta^2$, and hence $\xi^2\theta'$ is decreasing for $\xi>0$. 
The first interface and its radial flux
are therefore
\eq{
r_1=\frac{\xi_1}{\sqrt{a\Psi_c}},
\qquad
P=-r_1^2\Phi'(r_1)
=
\mu\sqrt{\frac{\Psi_c}{a}}.
}
Since \(\theta>0\) before its first zero, the negative core condition
\(\Phi>L\) is self-consistent.

\paragraph{Vacuum gap.}

For \(r_1<r<r_2\), both densities vanish, so \(\Delta\Phi=0\). Matching
\(\Phi\) and \(\Phi'\) at \(r=r_1\) gives
\eq{
\Phi(r)
=
L-P\left(\frac1{r_1}-\frac1r\right).
}
The positive shell begins at the first radius \(r_2>r_1\) such that
\(\Phi(r_2)=E_+\). Since \(\delta=L-E_+\), we obtain
\eq{
r_2
=
\left(\frac1{r_1}-\frac{\delta}{P}\right)^{-1}.
}
Thus, \(r_2\) is finite and satisfies \(r_2>r_1\) provided that
\eq{
\frac{P}{r_1}
=
\frac{\mu\Psi_c}{\xi_1}
>
\delta.
}
Moreover, \(\Phi\) decreases monotonically from \(L\) to \(E_+\) on
\(r_1<r<r_2\), so the vacuum gap condition \(E_+<\Phi<L\) is
self-consistent. 
For later convenience, set
\eq{
\tau
=
\frac{P}{r_1}-\delta
=
\frac{\mu\Psi_c}{\xi_1}-\delta
>0.
}
Then the matching relation $\Phi\lr{r_2}=E_+$ gives \(P/r_2=\tau\).

\paragraph{Positive shell.}

For \(r_2<r<R\), we have \(\Phi<E_+\). Hence, \(\rho_+>0\) and
\(\rho_-=0\). Set \(y(r)=E_+-\Phi(r)>0\). Then
\eq{
y''+\frac{2}{r}y'+ay^2=0,
}
with matching conditions \(y(r_2)=0\) and
\(y'(r_2)=P/r_2^2\). 
We now show that, for a suitable choice of \(\Psi_c\), the solution returns
to zero at a finite radius while satisfying the exterior vacuum condition.
Set
\eq{
s=\frac{P}{r_2^2},
\qquad
\ell=\lr{as}^{-1/3},
\qquad
\epsilon=\frac{\ell}{r_2}
=\lr{aPr_2}^{-1/3},
}
and introduce the stretched coordinate \(x\) and the rescaled profile
\(u_\epsilon\) by
\eq{
r=r_2+\ell x,
\qquad
y(r)=s\ell u_\epsilon(x).
} 
The positive shell problem becomes
\eq{
u_\epsilon''
+
\frac{2\epsilon}{1+\epsilon x}u_\epsilon'
+
u_\epsilon^2
=
0,
\qquad
u_\epsilon(0)=0,
\qquad
u_\epsilon'(0)=1.
}
Choose \(\Psi_c>\delta\xi_1/\mu\) and let it approach
\(\delta\xi_1/\mu\) from above. Then \(\tau\rightarrow0^+\),
\(r_2=P/\tau\rightarrow+\infty\), and
\(\epsilon=\lr{aP^2/\tau}^{-1/3}\rightarrow0\). In this limit,
\(u_\epsilon\) converges to the solution of
\eq{
u_0''+u_0^2=0,
\qquad
u_0(0)=0,
\qquad
u_0'(0)=1,
\qquad
\frac12\lr{u_0'}^2+\frac13u_0^3=\frac12.
}
The solution \(u_0\) returns to zero at a finite first positive point
\(x_0\), with \(u_0'(x_0)=-1\). Since this zero is transverse, continuous
dependence on \(\epsilon\) implies that, for sufficiently small
\(\epsilon>0\), \(u_\epsilon\) has a finite first positive zero
\(x_\epsilon\), with
\(x_\epsilon\rightarrow x_0\) and
\(u_\epsilon'(x_\epsilon)\rightarrow-1\). 
Hence the positive shell has a finite outer radius
\(R=r_2+\ell x_\epsilon\), with \(y(R)=0\) and \(\Phi(R)=E_+\).
Recalling \eqref{Phiext}, define \(Q=-R^2y'(R)>0\). Since
\(R/r_2=1+\epsilon x_\epsilon\rightarrow1\), we have
\eq{
\frac{Q/R}{P/r_2}
=
-\frac{R}{r_2}u_\epsilon'(x_\epsilon)
\longrightarrow1.
}
But \(P/r_2=\tau\rightarrow0\). Therefore \(Q/R\rightarrow0\), and,
for \(\Psi_c\) sufficiently close to \(\delta\xi_1/\mu\) from above, $
0<{Q}/{R}<\delta$.
Thus the positive shell terminates at a finite radius and the exterior
solution \eqref{Phiext} remains in the vacuum window
\(E_+\leq\Phi\leq L\).

\paragraph{Complete solution.}

\begin{figure}[h!]
    \centering
    \includegraphics[width=0.65\textwidth]{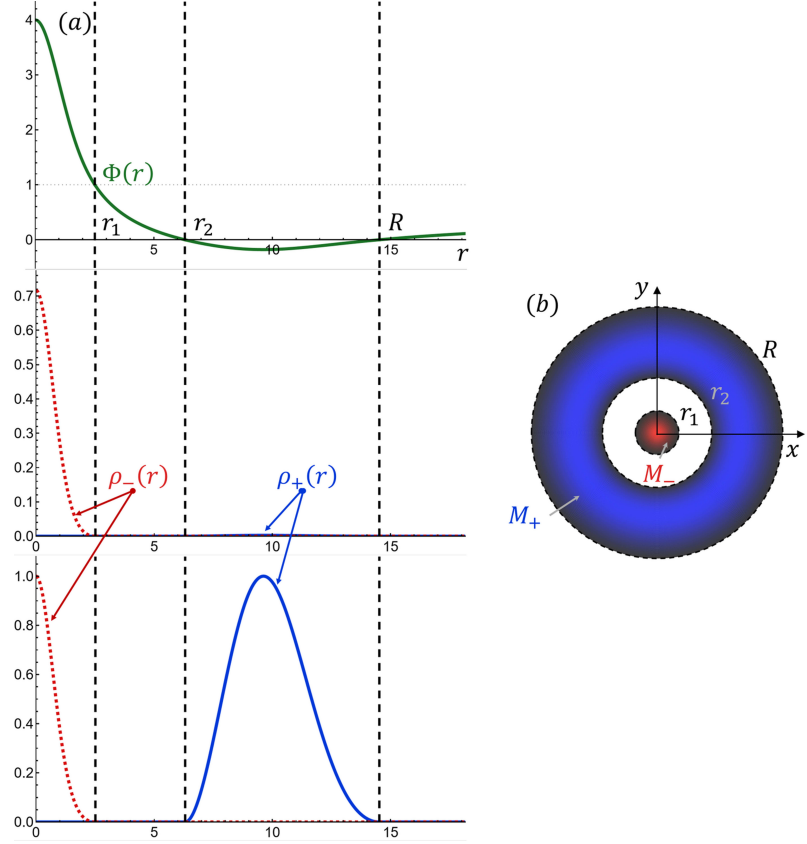}
    \caption{
    Gravitational-charge equilibrium for the \(n=1/2\) case.
    The parameters are \(a=b=1\), \(E_+=0\), \(E_-=-1\), so that
    \(L=1\) and \(\delta=1\), with central core amplitude
    \(\Psi_c=3\) and units chosen so that \(G=m=1\).
    The matching radii are
    \(r_1=2.51313\), \(r_2=6.31118\), and \(R=14.5254\), with exterior
    coefficient \(Q=8.16242\).
    Panel (a) shows the radial structure: the top subplot is the matched
    potential \(\Phi(r)\); the middle subplot shows the density profiles
    \(\rho_-(r)\) (red, dotted) and \(\rho_+(r)\) (blue, solid) on their
    absolute scales; the bottom subplot shows the same density profiles
    normalized by their respective maxima.
    The vertical dashed lines mark the interfaces \(r_1\), \(r_2\), and \(R\).
    Panel (b) is the corresponding \(x\)-\(y\) cross-section of the support:
    a negative mass core \(M_-\) occupies \(0\leq r\leq r_1\), a vacuum gap
    occupies \(r_1<r<r_2\), and a positive mass shell \(M_+\) occupies
    \(r_2\leq r\leq R\). The red and blue color intensities are normalized by the maxima of \(\rho_-\) and \(\rho_+\), respectively.
    }
    \label{fig4}
\end{figure}

Choose \(a=b>0\) and
\(\Psi_c>\delta\xi_1/\mu\) sufficiently close to
\(\delta\xi_1/\mu\) so that the preceding construction satisfies
\(Q/R<\delta\). Define
\[
r_1=\frac{\xi_1}{\sqrt{a\Psi_c}},
\qquad
P=\mu\sqrt{\frac{\Psi_c}{a}},
\qquad
r_2=
\left(\frac1{r_1}-\frac{\delta}{P}\right)^{-1}.
\]
Let \(y\) be the positive shell solution of
\(y''+2y'/r+ay^2=0\), with \(y(r_2)=0\) and
\(y'(r_2)=P/r_2^2\), let \(R>r_2\) be its first positive zero, and define
\(Q=-R^2y'(R)>0\). Then the full matched potential is
\begin{equation}\label{eq:Phi-piecewise-GVP-nhalf}
\Phi(r)
=
\begin{cases}
\displaystyle
L+\Psi_c\,\theta\lr{\sqrt{a\Psi_c}\,r},
& 0\leq r\leq r_1,
\\[2.0ex]
\displaystyle
L-P\left(\frac1{r_1}-\frac1r\right),
& r_1\leq r\leq r_2,
\\[2.0ex]
\displaystyle
E_+-y(r),
& r_2\leq r\leq R,
\\[2.0ex]
\displaystyle
E_+
+
Q\left(\frac1R-\frac1r\right),
& r\geq R.
\end{cases}
\end{equation}
With these definitions, \(\Phi\) and \(\Phi'\) are continuous at
\(r=r_1\), \(r=r_2\), and \(r=R\). Moreover,
\(Q=R^2\Phi'(R)=G\lr{M_+-M_-}>0\), and the active masses \(M_\pm\) are
finite because the solution is compactly supported. A plot of the solution
\eqref{eq:Phi-piecewise-GVP-nhalf} for a specific set of parameters is shown
in fig.~\ref{fig4}.

\begin{remark}
The restriction \(a=b\) is made to keep the construction and matching
calculations as simple as possible. The case \(a\neq b\) can be treated in
a similar fashion, but is not presented here.
\end{remark}

\section{Variations of the energy--Casimir functional}\label{sec:EC}

Stability of steady states for the standard gravitational Vlasov-Poisson system has been studied through the energy-Casimir method. 
Under suitable monotonicity, compactness, and regularity assumptions, nonlinear
stability results were obtained in
\cite{Rein1994EnergyCasimir,GuoRein1999}, while linear stability was analyzed
in \cite{BattMorrisonRein1995}. 

The purpose of this section is not to prove nonlinear stability of the
solutions constructed in Sec.~\ref{sec:compacteq}. Such a result would require
a carefully chosen class of equilibria and functional-analytic
estimates that go beyond the scope of the present paper, whose main aim is to
demonstrate the existence of compact Vlasov--Poisson equilibria with positive and negative mass.
Nevertheless, it is useful to determine the first and second variations of
the energy--Casimir functional. This calculation identifies the quadratic form
that would have to be controlled in a stability analysis.

Following \cite{HolmMarsdenRatiuWeinstein1985}, we first recall the notions of
formal nonlinear stability and nonlinear stability. Consider a dynamical system
\eq{
\frac{\p f}{\p t}=T\lrs{f},
}
where \(f\in X\), \(X\) is a Banach space, and \(T:X\to X\) is the evolution
operator. 

\begin{mydef}[Formal nonlinear stability]\label{def:fns}
Let \(\ml{E}\lrs{f}\) be a constant of motion. A steady
solution \(f^e\) such that $T\lrs{f^e}=0$ is formally nonlinearly stable with respect to \(\ml{E}\) if
\(f^e\) is a critical point
of \(\ml{E}\) and the second variation of \(\ml{E}\) at \(f^e\) is positive
definite on the admissible perturbation space:  
\eq{
\delta\ml{E}\lrs{f^e,h}=0,
\qquad
\delta^2\ml{E}\lrs{f^e,h}>0
\quad
\text{for all admissible perturbations}\quad h\ne0 .
}
\end{mydef}
Formal nonlinear stability, which is a key step in the energy--Casimir method, is not sufficient to establish nonlinear stability. The latter additionally requires the second variation to control the norm used to measure perturbations around the equilibrium.  

In general, the equilibrium \(f^e\) is nonlinearly stable if, for every
neighborhood \(U\) of \(f^e\), there exists a neighborhood \(V\) of \(f^e\) such
that every trajectory \(f(t)\) with \(f(0)\in V\) remains in \(U\) for all
\(t\ge0\).  In terms of a norm \(\Norm{\cdot}\) on $X$, we have: 

\begin{mydef}[Nonlinear stability]
The steady solution \(f^e\) is
nonlinearly stable if, for every \(\epsilon>0\), there exists \(\delta>0\) such
that
\eq{
\Norm{f(0)-f^e}<\delta
\qquad\Longrightarrow\qquad
\Norm{f(t)-f^e}<\epsilon
\quad
\text{for all }t\ge0 .
}
\end{mydef}

In the following, we examine the first and second variations of
the energy--Casimir functional \(\ml{E}\lrs{f_+,f_-}\) for the Bondi and
gravitational-charge Vlasov--Poisson equilibria. We focus, in particular, on
the structure of the second variation in relation to the notion of formal
nonlinear stability described in Def.~\ref{def:fns}. 

\subsection{Energy-Casimir variations in the Bondi sign  convention}

For the Vlasov--Poisson system~\ref{VP} with Bondi negative mass, the following energy
Hamiltonian functional is a constant of motion:
\eq{
\ml{H}\lrs{f_+,f_-}
=
\int_{\mathbb{R}^6}
\frac{\abs{\bol{v}}^2}{2}\lr{f_+-f_-}\,d\bol{x}d\bol{v}
+
\frac{1}{2}
\int_{\mathbb{R}^3}
\lr{\rho_+-\rho_-}\Phi\,d\bol{x}.
\label{HBondi}
}
The conservation of \eqref{HBondi} can be verified directly from system
\eqref{VP}, provided that the distribution functions \(f_{\pm}\), their
derivatives, and the associated potential have sufficient regularity and
decay at infinity to justify the required integrations by parts.
Similarly, 
functionals of the form
\eq{
\ml{N}
=
\sum_{\sigma=\pm}
\int_{\mathbb{R}^6}
N_{\sigma}\lr{f_{\sigma}}\,d\bol{x}d\bol{v},
\label{NBondi}
}
are constants of motion. Here, for each \(\sigma=\pm\),
\(N_{\sigma}:[0,\infty)\rightarrow\mathbb{R}\) is a sufficiently regular
function satisfying \(N_{\sigma}(0)=0\), so that the integral over the
exterior vacuum is finite. 
Next, we define the energy--Casimir functional
\eq{
\ml{E}=\ml{H}+\ml{N},\label{EC}
}
and consider admissible perturbations
\(\delta f_{\pm}=f_{\pm}-f_{\pm}^e\) around an equilibrium state
\(f_{\pm}^e\). Specifically, we restrict attention to support-preserving
perturbations, namely perturbations \(\delta f_\pm\) satisfying
\eq{
\operatorname{supp}\delta f_{\pm}
\subseteq
S_{\pm}
=
\operatorname{supp}f_{\pm}^e.\label{suppdelta}
}
This restriction is natural for the compact equilibria considered here, since
the perturbations then describe internal rearrangements of the particles
already present in the isolated configuration, rather than the creation of
particles in the exterior vacuum region. 
If one also wishes to fix the particle
numbers of each species, one may further impose
$\int_{\mathbb{R}^6}
\delta f_{\pm}\,d\bol{x}d\bol{v}
=0$. 
Within this fixed-support class, the cutoff relation defining \(f_\sigma^e\)
is invertible on \(S_\sigma\), and the Casimir functions can be chosen so that
the first variation of the energy--Casimir functional vanishes. Perturbations
that change the support, or populate the exterior vacuum, involve an additional
free-boundary issue and are not considered here.

Writing
\(\rho_{\pm}=\rho_{\pm}^e+\delta\rho_{\pm}\) and
\(\Phi=\Phi^e+\delta\Phi\), with
\(\Delta\delta\Phi=4\pi Gm\lr{\delta\rho_+-\delta\rho_-}\), the total change
in the Hamiltonian \(\ml{H}\) caused by the perturbations is
\eq{
\ml{H}\lrs{f^e_++\delta f_+,f^e_-+\delta f_-}
-
\ml{H}\lrs{f^e_+,f_-^e}
=
\int_{\mathbb{R}^6}
\lr{\delta f_+-\delta f_-}
\lr{\frac{\abs{\bol{v}}^2}{2}+\Phi^e}\,
d\bol{x}d\bol{v}
-
\frac{1}{8\pi Gm}
\int_{\mathbb{R}^3}
\abs{\nabla\delta\Phi}^2\,d\bol{x}.
}
It follows that the variations of the energy--Casimir functional evaluated at the equilibrium point are
\sys{
&\delta\ml{E}^e
=
\int_{\mathbb{R}^6}
\lr{\delta f_+-\delta f_-}
\lr{\frac{\abs{\bol{v}}^2}{2}+\Phi^e}\,
d\bol{x}d\bol{v}
+
\sum_{\sigma=\pm}
\int_{\mathbb{R}^6}
N_{\sigma}'\lr{f_{\sigma}^e}
\delta f_{\sigma}\,d\bol{x}d\bol{v},
\\
&\delta^2\ml{E}^e
=
-\frac{1}{4\pi Gm}
\int_{\mathbb{R}^3}
\abs{\nabla\delta\Phi}^2\,d\bol{x} 
+
\sum_{\sigma=\pm}
\int_{\mathbb{R}^6}
N_{\sigma}''\lr{f_{\sigma}^e}
\delta f_{\sigma}^2\,d\bol{x}d\bol{v},
}{Evars}
where \(N_{\sigma}'=dN_{\sigma}/df_{\sigma}\) and
\(N_{\sigma}''=d^2N_{\sigma}/df_{\sigma}^2\).

The condition that \(f_{\pm}^e\) is a critical point of the energy--Casimir
functional \(\ml{E}\), within the support-preserving perturbation class, can be
satisfied by demanding that
\eq{
N_{\sigma}'\lr{f_{\sigma}^e}
=
-\sigma
\lr{
\frac{\abs{\bol{v}}^2}{2}+\Phi^e
}
\qquad
\text{on }S_\sigma .
}
For \(n_\sigma=-1/2\), we obtain
\eq{
N_{\sigma}\lr{f_\sigma}
=
\sigma
\lr{
-\frac{A_{\sigma}^2}{f_{\sigma}}
-
E_{\sigma}f_{\sigma}}
+
c_{\sigma}^{-1/2}
,
\qquad
c_{\sigma}^{-1/2}\in\mathbb{R},
\qquad
\sigma=\pm .
}
Similarly, for \(n_{\sigma}=1/2\), we obtain
\eq{
N_{\sigma}\lr{f_\sigma}
=
\sigma
\lr{
\frac{f_{\sigma}^3}{3A_{\sigma}^2}
-
E_{\sigma}f_{\sigma}}
+
c_{\sigma}^{1/2}
,
\qquad
c_{\sigma}^{1/2}\in\mathbb{R},
\qquad
\sigma=\pm .
}
For the second variation of the \(n_\sigma=-1/2\) case, we thus find
\eq{
\delta^2\ml{E}^e
=&
\frac{2}{A_-}
\int_{S_-}
\lr{
E_-
-
\frac{\abs{\bol{v}}^2}{2}
-
\Phi^e
}^{3/2}
\delta f_-^2\,d\bol{x}d\bol{v}
-
\frac{2}{A_+}
\int_{S_+}
\lr{
E_+
-
\frac{\abs{\bol{v}}^2}{2}
-
\Phi^e
}^{3/2}
\delta f_+^2\,d\bol{x}d\bol{v}
\\&-
\frac{1}{4\pi Gm}
\int_{\mathbb{R}^3}
\abs{\nabla\delta\Phi}^2\,d\bol{x}.
}
When \(n_{\sigma}=1/2\), we have
\eq{
\delta^2\ml{E}^e
=&
\frac{2}{A_+}
\int_{S_+}
\sqrt{
E_+
-
\frac{\abs{\bol{v}}^2}{2}
-
\Phi^e
}\,
\delta f_{+}^2\,d\bol{x}d\bol{v}
-
\frac{2}{A_-}
\int_{S_-}
\sqrt{
E_-
-
\frac{\abs{\bol{v}}^2}{2}
-
\Phi^e
}\,
\delta f_{-}^2\,d\bol{x}d\bol{v}
\\&-
\frac{1}{4\pi Gm}
\int_{\mathbb{R}^3}
\abs{\nabla\delta\Phi}^2\,d\bol{x}.
}
We therefore see that, for the compactly supported Bondi negative mass
equilibria, the second variation is not sign definite a priori. In each of the
two cutoff cases above, one species contributes with a favorable sign, while
the other species and the field perturbation contribute with an unfavorable
sign. Thus formal nonlinear stability in this energy--Casimir sense would
require a choice of parameters and an admissible perturbation class for which
the favorable contribution controls the unfavorable terms.

\subsection{Energy-Casimir variations in the  gravitational-charge sign convention}

We now examine the first and second variations of the energy--Casimir
functional for the Vlasov--Poisson system \eqref{VP} with gravitational-charge negative mass.
Under the hypothesis that the distribution functions and their derivatives
vanish sufficiently rapidly at infinity, the conserved Hamiltonian functional
is 
\eq{
\ml{H}\lrs{f_+,f_-}
=
\int_{\mathbb{R}^6}
\frac{\abs{\bol{v}}^2}{2}\lr{f_++f_-}\,d\bol{x}d\bol{v}
+
\frac{1}{2}
\int_{\mathbb{R}^3}
\lr{\rho_+-\rho_-}\Phi\,d\bol{x}.
\label{Hgc}
}
The functionals \eqref{NBondi} 
are also constants of motion, 
and the energy--Casimir functional
takes the usual form $\ml{E}=\ml{H}+\ml{N}$. 
As in the Bondi case \eqref{suppdelta}, we consider support-preserving
perturbations \(\delta f_{\pm}=f_{\pm}-f_{\pm}^e\) around an equilibrium state
\(f_{\pm}^e\). Writing
\(\rho_{\pm}=\rho_{\pm}^e+\delta\rho_{\pm}\) and
\(\Phi=\Phi^e+\delta\Phi\), with
\(\Delta\delta\Phi=4\pi Gm\lr{\delta\rho_+-\delta\rho_-}\), the total change
in the Hamiltonian \eqref{Hgc} can be evaluated as
\eq{
\ml{H}\lrs{f^e_++\delta f_+,f^e_-+\delta f_-}
-
\ml{H}\lrs{f^e_+,f_-^e}
=&
\int_{\mathbb{R}^6}
\lrs{
\lr{\frac{\abs{\bol{v}}^2}{2}+\Phi^e}\delta f_+
+
\lr{\frac{\abs{\bol{v}}^2}{2}-\Phi^e}\delta f_-
}
\,d\bol{x}d\bol{v}
\\
&-
\frac{1}{8\pi Gm}
\int_{\mathbb{R}^3}
\abs{\nabla\delta\Phi}^2\,d\bol{x}.
}
It follows that the variations of the energy--Casimir functional are
\sys{
&\delta\ml{E}^e
=
\int_{\mathbb{R}^6}
\lrs{
\lr{\frac{\abs{\bol{v}}^2}{2}+\Phi^e+N_+'\lr{f_+^e}}\delta f_+
+
\lr{\frac{\abs{\bol{v}}^2}{2}-\Phi^e+N_-'\lr{f_-^e}}\delta f_-
}
\,d\bol{x}d\bol{v},
\\
&\delta^2\ml{E}^e
=
-
\frac{1}{4\pi Gm}
\int_{\mathbb{R}^3}
\abs{\nabla\delta\Phi}^2\,d\bol{x}
+
\sum_{\sigma=\pm}
\int_{S_\sigma}
N_{\sigma}''\lr{f_{\sigma}^e}
\delta f_{\sigma}^2\,d\bol{x}d\bol{v}.
}
{ECgcvars}
For the gravitational-charge--Vlasov--Poisson equilibria, the
criticality condition \(\delta\ml{E}^e=0\) within the support-preserving
perturbation class follows by demanding that
\eq{
N_{\sigma}'\lr{f_{\sigma}^e}=
-\frac{\abs{\bol{v}}^2}{2}-\sigma\Phi^e
\qquad {\rm on}~~S_{\sigma}.
}
When \(n_\sigma=-1/2\), we find
\eq{
N_{\sigma}\lr{f_\sigma}
=
-\frac{A_\sigma^2}{f_{\sigma}}
-
E_{\sigma}f_\sigma
+
c_\sigma^{-1/2},
\qquad
c_{\sigma}^{-1/2}\in\mathbb{R},
\qquad
\sigma=\pm .
}
For \(n_\sigma=1/2\), we obtain
\eq{
N_{\sigma}\lr{f_\sigma}
=
\frac{f_{\sigma}^3}{3A_{\sigma}^2}
-
E_{\sigma}f_{\sigma}
+
c_\sigma^{1/2},
\qquad
c_{\sigma}^{1/2}\in\mathbb{R},
\qquad
\sigma=\pm .
}
With these choices of the functions \(N_{\sigma}\), it remains to evaluate the
second variation at the equilibrium states. For \(n_{\sigma}=-1/2\), we have
\eq{
\delta^2\ml{E}^e
=
-
2\sum_{\sigma=\pm}
\int_{S_\sigma}
\frac{A_{\sigma}^2}{\lr{f_{\sigma}^e}^3}
\delta f_{\sigma}^2\,d\bol{x}d\bol{v}
-
\frac{1}{4\pi Gm}
\int_{\mathbb{R}^3}
\abs{\nabla\delta\Phi}^2\,d\bol{x}
\leq 0 .
\label{gcsecvarm}
}
Thus, for \(n_\sigma=-1/2\), the second variation is non-positive.
Equivalently
\(-\mathcal E\) has positive-definite second variation for any nontrivial perturbation. This is
compatible with formal stability in the fixed-support
energy--Casimir sense. 

For \(n_{\sigma}=1/2\), we have
\eq{
\delta^2\ml{E}^e
=
2\sum_{\sigma=\pm}
\int_{S_\sigma}
\frac{f_{\sigma}^e}{A_{\sigma}^2}
\delta f_{\sigma}^2\,d\bol{x}d\bol{v}
-
\frac{1}{4\pi Gm}
\int_{\mathbb{R}^3}
\abs{\nabla\delta\Phi}^2\,d\bol{x}.
\label{gcsecvarp}
}
In this case, the second variation consists of two non-negative microscopic
contributions and a non-positive field contribution. Hence formal nonlinear
stability in this fixed-support energy--Casimir sense rests on the possibility
of identifying a parameter regime and an admissible perturbation class for
which the positive microscopic terms dominate the negative field term.

\section{A geometric interpretation of the Bondi sign convention}
\label{sec:oriented-mass}

In this final section, we briefly discuss a possible geometric interpretation
of the sign appearing in the Bondi negative mass convention. Let \(\Omega\) be a four-dimensional smooth Lorentzian manifold with metric
\(g_{\mu\nu}\), and consider the trivial double cover
\eq{
\widehat{\Omega}
=
\Omega\times\lrc{+1,-1}.
}
The two sheets
\(\Omega\times\lrc{+1}\) and \(\Omega\times\lrc{-1}\)
represent two distinct matter sectors over the same spacetime base
\(\Omega\). A matter element is therefore described by a pair
\(\lr{x,\sigma}\), where \(x\in\Omega\) and \(\sigma=\pm1\) labels the
sheet. We regard \(g_{\mu\nu}\) as a common gravitational field on
\(\Omega\), whose pullback determines the spacetime geometry experienced
by matter on both sheets.

Recall that, in Einstein's field equations, mass enters through the
stress-energy tensor on the right-hand side, with no independent
distinction between inertial, passive gravitational, and active gravitational
mass. 
In the weak-field limit, one writes
\eq{
g_{00}
=
-\lr{1+\frac{2\Phi}{c^2}}
+
O\lr{c^{-4}},
}
so that \(\Phi/c^2\) measures the deviation of the time  component of the
metric from its Minkowski value. The Newtonian limit is obtained under the
condition \(\abs{\Phi}/c^2\ll1\), and the classical Poisson equation follows
from Einstein's field equations in this limit; see, for example,
\cite{FrankelGC}.

We now postulate that the sheet label \(\sigma\) determines the sign with
which each matter sector sources the common gravitational field.

\begin{mydef}[Signed mass]
On the trivial double cover
\(\widehat{\Omega}=\Omega\times\lrc{+1,-1}\), a signed mass is a pair
\(\lr{m,\sigma}\), where \(m>0\) is the absolute mass and
\(\sigma=\pm1\) is the sheet label. The associated signed gravitational
mass is defined by $m_{\sigma}=\sigma m$.
\end{mydef}

Thus matter on the \(\sigma=1\) sheet behaves as ordinary positive mass,
while matter on the \(\sigma=-1\) sheet contributes to the gravitational
source with the opposite sign. Let \(T_{\sigma}^{\mu\nu}\) denote the
ordinary, unsigned stress-energy tensor associated with the matter sector
\(\sigma\), regarded as a tensor field on \(\Omega\) through the 
identification of each sheet with the base spacetime. For example, if the
\(\sigma=-1\) sector is a pressureless fluid with positive rest-mass density
\(m\rho_-\) and four-velocity \(u_-^\mu\), then
\(T_-^{\mu\nu}=m\rho_-u_-^\mu u_-^\nu\), while its contribution to the
gravitational source is \(-T_-^{\mu\nu}\). 
With this convention, we postulate the signed Einstein equations
\eq{
R^{\mu\nu}
-
\frac{1}{2}g^{\mu\nu}R
=
\frac{8\pi G}{c^4}
\sum_{\sigma=\pm}
\sigma T_{\sigma}^{\mu\nu}.
}
When \(T_-^{\mu\nu}=0\), one recovers the usual Einstein equations for
ordinary positive mass. When \(T_+^{\mu\nu}=0\) and
\(T_-^{\mu\nu}\neq0\), the sign of the gravitational source is reversed
relative to the ordinary matter configuration with the same unsigned
stress-energy tensor.

In the weak-field, slowly moving regime, we have
\(T_{\sigma}^{00}\simeq m\rho_{\sigma}c^2\), where
\(\rho_{\sigma}\geq0\) denotes the unsigned number density on the
\(\sigma\)-sheet. The signed Einstein equations therefore reduce to
\eq{
\Delta\Phi
=
4\pi Gm \lr{\rho_+-\rho_-}.
}
At the same time, matter on both sheets moves according to the geodesic
equation of the same metric \(g_{\mu\nu}\), and hence obeys the same
Newtonian acceleration law 
$\ddot{\bol{x}}
=
-\nabla\Phi$.
Thus, in the weak-field limit, matter on the \(\sigma=-1\) sheet sources
the potential with the opposite sign but falls in the same gravitational
field as matter on the \(\sigma=1\) sheet. This reproduces the weak-field Vlasov--Poisson equations associated with the Bondi sign convention. 

\begin{remark}[Interactions between the two matter sectors]
The two sheets
\(\Omega\times\lrc{+1}\) and \(\Omega\times\lrc{-1}\)
are disconnected components of \(\widehat{\Omega}\); in particular, no
continuous path in \(\widehat{\Omega}\) connects points belonging to
different sheets. Hence, if non-gravitational interactions are mediated by
local fields propagating on \(\widehat{\Omega}\), their mediators cannot
propagate from one sheet to the other, and the two matter sectors cannot
interact through such forces. They nevertheless interact gravitationally
through the common metric on \(\Omega\), whose source is the signed
stress-energy tensor \(T_+^{\mu\nu}-T_-^{\mu\nu}\).
\end{remark} 

  
\section*{Data availability}
Data sharing not applicable to this article as no datasets were generated or analysed during the current study.





\bibliographystyle{alphaurl}
\bibliography{refs_initials}



 



\end{document}